\documentclass[lettersize,journal]{IEEEtran}
\usepackage{amsmath,amsfonts}
\usepackage{algorithmic}
\usepackage{algorithm}
\usepackage{array}
\usepackage{textcomp}
\usepackage{stfloats}
\usepackage{url}
\usepackage{verbatim}
\usepackage{graphicx}
\usepackage{cite}
\usepackage{hyperref}
\usepackage{xcolor}
\usepackage{tabularx}
\usepackage{longtable}
\usepackage{multirow}
\usepackage{enumitem}
\usepackage{threeparttable} 
\usepackage{booktabs}
\usepackage{wasysym} 
\usepackage{amssymb} 
\usepackage{pifont}
\usepackage{xurl}
\usepackage{subcaption}

\newcommand{\high}{\CIRCLE}
\newcommand{\medium}{\LEFTcircle}
\newcommand{\low}{\Circle}

\usepackage{tikz}

\begin{document}

\title{A Comprehensive Survey of Advanced Persistent Threat Attribution: Taxonomy, Methods, Challenges and Open Research Problems}

\author{Nanda Rani, Bikash Saha, Sandeep Kumar Shukla ~\IEEEmembership{}
\thanks{Nanda Rani, Bikash Saha, and Sandeep Kumar Shukla are with Department of Computer Science \& Engineering, Indian Institute of Technology Kanpur, India (email: \href{mailto:nandarani@cse.iitk.ac.in}{nandarani@cse.iitk.ac.in}; \href{mailto:bikash@cse.iitk.ac.in}{bikash@cse.iitk.ac.in}; \href{mailto:sandeeps@cse.iitk.ac.in}{sandeeps@cse.iitk.ac.in})}
}



\maketitle

\begin{abstract}
Advanced Persistent Threat (APT) attribution is a critical challenge in cybersecurity and implies the process of accurately identifying the perpetrators behind sophisticated cyber attacks. It can significantly enhance defense mechanisms and inform strategic responses. 
With the growing prominence of artificial intelligence (AI) and machine learning (ML) techniques, researchers are increasingly focused on developing automated solutions to link cyber threats to responsible actors, moving away from traditional manual methods. 
Previous literature on automated threat attribution lacks a systematic review of automated methods and relevant artifacts that can aid in the attribution process.
To address these gaps and provide context on the current state of threat attribution, we present a comprehensive survey of automated APT attribution.
The presented survey starts with understanding the dispersed artifacts and provides a comprehensive taxonomy of the artifacts that aid in attribution. We comprehensively review and present the classification of the available attribution datasets and current automated APT attribution methods. Further, we raise critical comments on current literature methods, discuss challenges in automated attribution, and direct toward open research problems. 
This survey reveals significant opportunities for future research in APT attribution to address current gaps and challenges.
By identifying strengths and limitations in current practices, this survey provides a foundation for future research and development in automated, reliable, and actionable APT attribution methods.
\end{abstract}


\section{Introduction}
\label{sec:introduction}

When firewalls and antivirus softwares become mere hurdles rather than barriers, we enter the realm of Advanced Persistent Threats (APTs). 
APT attacks are attack campaigns orchestrated by highly organized and often state-sponsored threat groups that operate covertly and methodically over prolonged periods~\cite{ahmad2019strategically}. 
APTs set themselves apart from conventional cyber-attacks by their stealthiness, persistence, and precision in targeting~\cite{chen2014study,alshamrani2019survey}. Unlike typical cyber-attacks, which aim for immediate and indiscriminate disruption or data exfiltration, APTs are orchestrated by resourceful adversaries who carefully conduct multi-phase operations. These adversaries excel at evading detection over prolonged periods, continuously adapting to security measures. APTs are recognized for their underlying intent and the sophisticated attack methodology~\cite{alshamrani2019survey}. APTs employ meticulously crafted tactics, techniques, and procedures (TTPs) tailored to breach specific targets or organization, in contrast to traditional attacks that often rely on automated tools with widespread and indiscriminate targets. This high degree of customization and persistence makes APTs exceptionally difficult to detect and counteract.


The intricate and sustained nature of APTs demands a deep understanding of the actors orchestrating these campaigns, along with effective detection mechanisms. This deeper analysis is essential to anticipate future moves and tailor defenses accordingly.
The process of identifying and linking sophisticated attacks back to the associated and responsible threat actors or nation-states is known as APT Attribution~\cite{steffens2020attribution,skopik2020under,sharma2023advanced,mei2022review}. It is vital for understanding the complex nature of these threats and the actors behind them, providing crucial insights into the broader context of cyber activities on a global scale. Threat attribution remains a cornerstone in cybersecurity, as it helps to reveal the intricate web of motives and strategies employed by these malicious entities.

Attributing APT attacks is instrumental in understanding the attacker's motivations, strategies, and techniques. Further, such understanding, in turn, informs the development of targeted defensive strategy, proactive countermeasures and enhances the resilience of potential targets~\cite{singer2014cybersecurity}.
Beyond bolstering cybersecurity, precise attribution is the groundwork for legal and diplomatic actions. It enables victims to seek justice through legal channels, impose sanctions, or engage in diplomatic efforts to curb malicious activities~\cite{steffens2020attribution}. 
In recent times, the FBI (Federal Bureau of Investigation) charged Lazarus group~\cite{MITRE_G0032} linked three individuals: Park Jin Hyok~\cite{FBI_ParkJinHyok}, Kim Il~\cite{FBI_KimIl}, and Jon Chang Hyok~\cite{FBI_JonChangHyok}.
On a broader scale, accurate and timely attribution may act as a deterrent. It also signals to potential aggressors that their actions will not remain anonymous, and they will be held accountable. Further, attribution facilitates the sharing of actionable threat intelligence across organizations and sectors, which further improves the overall cybersecurity posture of the organization~\cite{kaspersky_whitepaper}. Thus, threat attribution is crucial in immediate response and long-term security planning.


\begin{table*}[]
    \centering
    \caption{Comparison with related existing survey}
    \label{tab:surveycomparison}
    \begin{threeparttable}
    \begin{tabularx}{\textwidth}{p{1.3cm}p{1cm}XXXXXXp{1.9cm}}
     \hline
     \textbf{Paper} & \textbf{Year} & \textbf{Attribution Evolution} & \textbf{Attribution Case Study} & \textbf{Artifact Taxonomy} & \textbf{Method Classification} & \textbf{Dataset Classification} & \textbf{Challenges} & \textbf{Open Research Problems}  \\
     \hline
     \cite{mei2022review} & $2022$ & \quad \ding{55} & \quad \ding{55} & \quad \ding{55} & \quad \ding{51} & \quad \ding{55} & \quad \ding{55} & \quad \ding{55} \\
     \cite{tang2022advanced} & $2022$ & \quad \ding{55} & \quad \ding{55} & \quad \ding{55} & \quad \ding{51} & \quad \ding{55} & \quad \ding{51}  & \quad \ding{55} \\
     \cite{sharma2023advanced} & $2023$ & \quad \ding{55} &  \quad \ding{51} &  \quad \ding{55} & \quad \ding{55} & \quad \ding{55} & \quad \ding{55} & \quad \ding{55} \\
     \cite{gray2024identifying}\tnote{ *} & $2024$ & \quad \ding{55} & \quad \ding{55} & \quad \ding{55} & \quad \ding{51} & \quad \ding{51} & \quad \ding{51} & \quad \ding{51}\\
     Our Survey & $2024$ & \quad \ding{51} &  \quad \ding{51}  & \quad \ding{51}  &  \quad \ding{51} & \quad \ding{51} & \quad \ding{51} & \quad \ding{51} \\
\end{tabularx}
    \begin{tablenotes}
        \item[*] This survey is specific to malware-based attribution methods only
        \end{tablenotes}
    \end{threeparttable}
\end{table*}

Attributing attacks involves a multifaceted approach, combining digital forensics, intelligence analysis, and geopolitical insights. These approaches peel back the layers of anonymity or evasiveness that attackers hide behind~\cite{lin2016attribution,knake2010untangling}. This intricate process employs a blend of technical and non-technical artifacts to identify the unique digital fingerprints left by attackers~\cite{brandao2021advanced}. Threat analyst teams scrutinize these indicators, comparing them against vast databases of known threats. It further aids in finding similarities or establishing new profiles for unidentified attackers. Analysts perform this analysis on traditional frameworks such as Diamond Model~\cite{caltagirone2013diamond}, CAM~\cite{skopik2020under}, Q model~\cite{rid2015attributing}, triangle model~\cite{warikoo2021triangle}, and MICTIC framework~\cite{steffens2020attribution}, which facilitate multi-dimension approach to connect the dots between collected artifacts and suspected threat actor. While traditional manual methods allow for deep and contextual analysis of attacks, their effectiveness and scalability are increasingly challenged by the rapidly evolving cyber threat landscape~\cite{mei2022review,rid2015attributing}.
The advent of advanced automated technologies such as artificial intelligence (AI) and machine learning (ML) presents an opportunity to enhance APT attribution's performance, efficiency, reliability, and scalability~\cite{avellaneda2019using}. Further, the literature of automated threat attribution leverages series of attack artifacts with state-of-the-art automated technological capabilities to uncover the threat actors. We review such available automated threat attribution methods and present a comprehensive survey of automated APT attribution. 

\subsection{Motivation}
\label{subsec:motivation}
The motivation behind this survey on automated APT attribution stems from the growing frequency and sophistication of APT attacks, which necessitates scalable and efficient attribution methods. Advanced technologies such as AI and ML offer opportunities to enhance the performance, reliability, and scalability of these methods. Despite the availability of several automated approaches in the literature, they are dispersed and lack a systematic synthesis. This survey aims to consolidate and critically analyze these methods, highlighting strengths, potential limitations and understanding their real-world applications. The goal is to advance the field by identifying effective strategies for automated APT attribution and addressing remaining gaps and challenges.

\subsection{Review of prior surveys and gap analysis}
\label{subsec:reviewofliterature}
The current surveys and research papers have explored various aspects of APTs and attribution techniques. Previous surveys have predominantly focused on traditional attribution methods~\cite{mei2022review,sharma2023advanced}, the detection and mitigation of APTs~\cite{alshamrani2019survey,sharma2023advanced,quintero2020new,buchta2024advanced}, profiling techniques~\cite{tang2022advanced}, threat intelligence of APTs~\cite{schlette2021comparative,wagner2019cyber,sun2023cyber}, and threat modelling of APTs~\cite{tatam2021review,bodeau2018cyber}. While these surveys provide valuable insights about APTs, there is a notable gap in the literature concerning the comprehensive review of automated APT attribution methods. 
The literature on the APT attribution methods is dispersed and scattered. In our exploration, we find very few surveys present in the literature~\cite{mei2022review,tang2022advanced,sharma2023advanced,gray2024identifying}, which discuss threat attribution, but that too lacks detailed discussion, comparison of available attribution methods, discussion of current attribution challenges and open research problems.
There is also a lack of systematic literature that can summarize the complete flow and end-to-end automated attribution process. Such summarization facilitates a broader understanding of the community. Therefore, this paper systematizes the automated APT attribution methods available in the literature. The comprehensive survey aids in understanding the available effective automated attribution methodology. 
Table~\ref{tab:surveycomparison} compares this survey with the related existing study.

\subsection{Scope and contribution}
\label{subsec:scoeandobjective}

This survey consolidates current knowledge, highlights recent advancements, and identifies gaps in existing methodologies. It also serves as a bridge between various isolated studies and offers a unified perspective that is currently lacking.
This consolidation is crucial for advancing research in this domain and provides cybersecurity practitioners and policymakers with a comprehensive overview of effective methods. This survey thoroughly reviews the state-of-the-art techniques for automated APT attribution by presenting a detailed case study and assessing the effectiveness of various automated methods.
The contributions of this paper include review of available attribution dataset and methods, a critical analysis of current methodologies, identification of existing gaps and challenges, and presenting the open research problem toward effective and reliable automated APT attribution. This survey thus serves as a valuable resource for researchers, blue and red teams of the organizations and practitioners in cybersecurity. It also offers insights and guidance for advancing the practice of automated APT attribution.
The key contributions of this survey are the following:
\begin{enumerate}[label=(\alph*)]
    \item We contextualize the evolution of APT attribution by classifying the advancements in the attribution methods since beginning.
    \item  We taxonomize the dispersed attribution artifacts, which can be utilized effectively in the attribution process.
    \item We systematically illustrate the end-to-end attribution process, detailing each phase from data collection to analysis. 
    \item We explore and classify all (structured and unstructured) publicly available threat data, which can be leveraged to develop a dataset for presenting new attribution methodologies.
    \item We classify the current literature on attribution methodologies and comprehensively compare these methodologies based on various bases.
    \item We examine the current technical and methodological challenges in automated attribution that hinder accurate and reliable threat attribution.
    \item We identify and propose potential open research problems in automated attribution, suggesting novel problems and areas where further investigation and innovation can significantly advance automated threat attribution.
\end{enumerate}

\subsection{Structure of the Paper}
\label{subsec:structure}
\begin{figure}[!h]
    \centering
    \includegraphics[width=\columnwidth]{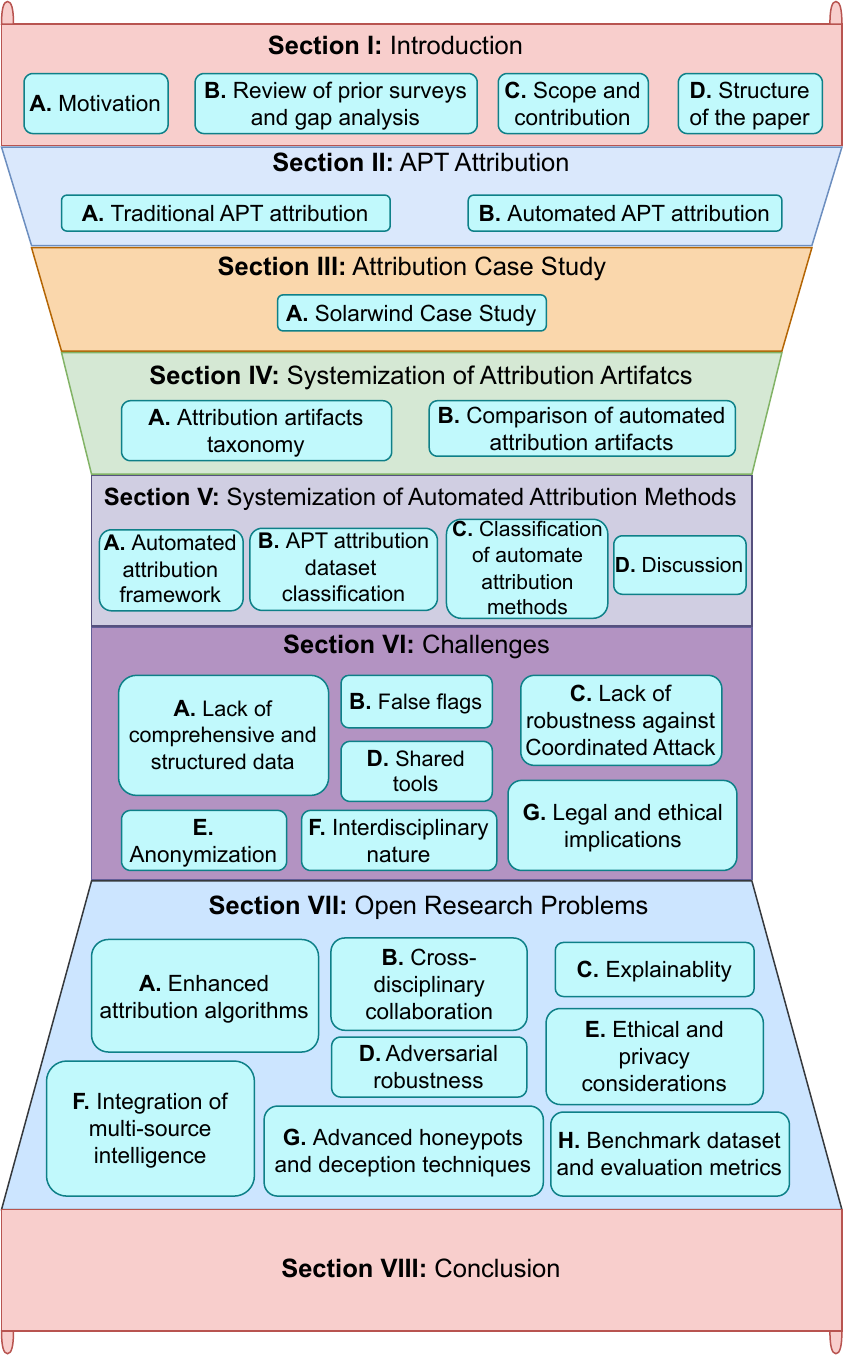}
    \caption{Organization of this paper}
    \label{fig:structure}
\end{figure}

The structure of this paper is shown in Fig~\ref{fig:structure}. Section~\ref{sec:aptattirbution} presents the evolution and current state of APT attribution, and Section~\ref{sec:casestudy} details an attribution case study of past attack campaigns. Then, the proposed attribution artifact taxonomy is presented in Section~\ref{sec:systemizationattributionartifacts}, and a survey of the current attribution datasets and methods is discussed in Section~\ref{sec:systemizationattributionmethods}. Further, we present the current attribution challenges in Section~\ref{sec:challenges} and discuss open research problems and future research directions in Section~\ref{sec:openresearchproblems}. Section~\ref{sec:conclusion} concludes the presented survey. The list of acronyms used in this paper is listed in Table~\ref{tab:acronyms}.

\begin{table}[!h]
    \centering
    \caption{List of acronyms and their corresponding descriptions}
    \begin{tabular}{|c|c|}
    \hline
     \textbf{Acronym} & \textbf{Description}  \\
     \hline
     AI & Artificial Intelligence \\
     API & Application Programming Interface \\
     APT & Advanced Persistent Threat \\
     ATT\&CK & Adversarial Tactics, Techniques and Common Knowledge  \\
     C\&C & Command and Control \\
     CFG & Control Flow Graph \\
     CTI & Cyber Threat Intelligence \\
     CVE & Common Vulnerabilities and Exposure \\
     DNN & Deep Neural Network \\
     DNS & Domain Name System \\
     DT & Decision Tree \\
     GNN & Graph Neural Network \\
     IoC & Indicators of Compromise \\
     KNN & K-nearest Neighbor \\
     LightGBM & Light Gradient Boosting Machine \\
     ML & Machine Learning \\
     NB & Naïve Bayes \\
     OSINT & Open Source INTelligence  \\
     RF & Random Forest \\
     SVM & Support Vector Machine \\
     TF-IDF & Term Frequency-Inverse Document Frequency \\
     TTP & Tactics, Techniques, and Procedures \\
     XGBoost & Extreme Gradient Boosting  \\
     \hline
\end{tabular}
    \label{tab:acronyms}
\end{table}

\section{APT Attribution}
\label{sec:aptattirbution}


Attribution in the context of cybersecurity, particularly concerning APTs, is distinct from traditional cyber criminal investigations. Attribution involves tracing the source of cyber attacks, typically tied to state-sponsored entities. 
The evolution of cyber threat attribution methodologies has seen significant advancements over the decades. 

In the early beginnings ($1986$-$2000$), manual methods prevail, exemplified by the "Cuckoo's Egg" incident~\cite{mei2022review,stoll2005cuckoo}, where digital forensics plays a crucial role in tracking cyber intrusions. During this period, the U.S. military also introduces the concept of attribution~\cite{mei2022review}, laying the groundwork for systematic approaches. 
The goal was to identify the actors behind cyberattacks, particularly nation-states and organized cybercriminal groups. Initial attribution efforts were hindered by the lack of standardized methodologies and the complexity of APT attack vectors. As cyber incidents grew more complex and prevalent, the need for structured approaches became apparent, leading to the development of foundational frameworks and methodologies. 

From $2000$ to $2015$, a more systematic approach emerges to enhance coordination and the development of frameworks such as the Diamond Model~\cite{caltagirone2013diamond} and MITRE ATT\&CK~\cite{mitreattack}, which provide structured methods for documenting adversary tactics, techniques, and procedures. The documented adversary's modus operandi are correlated to identify any link between attack incidents with historical attacks. 

With the emergence of automation and ML in analyzing vast datasets quickly and efficiently, the period from $2015$ to $2023$ witnesses the rise of automated approaches along with behavioral analytics and enhanced threat intelligence sharing to improve attribution accuracy. 

\begin{figure*}[!ht]
    \centering
    \includegraphics[width=\linewidth]{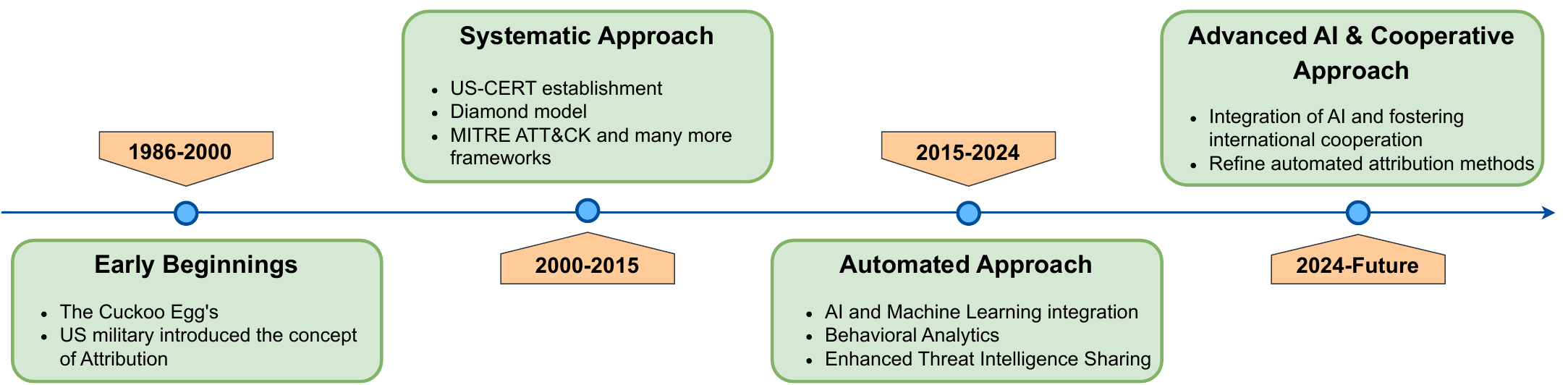}
    \caption{Evolution of cyber threat attribution}
    \label{fig:attributionevolution}
\end{figure*}

Looking forward, the advanced AI and cooperative approach ($2024$-future) emphasize the continued integration of AI and fostering international cooperation, aiming to refine automated attribution methods and enhance their robustness against evolving cyber threats~\cite{evolution_of_cyber_attribution}. A timeline of evolution of attribution methodologies is shown in Fig.~\ref{fig:attributionevolution}.

\subsection{Traditional APT attribution}
\label{subsec:traditionalattribtuion}

With the evolution of cyber threats, the techniques for attribution have also advanced towards structure-based frameworks. In the period of $2000$ to $2015$, the cybersecurity community or researchers shift their focus to develop more systematic approaches. During this period, we observed the rise of frameworks such as the Lockheed Martin Cyber Kill Chain (CKC)~\cite{lockheed2011cyber}, MITRE ATT\&CK framework~\cite{mitreattack}, Diamond Model~\cite{caltagirone2013diamond} and MICTIC framework~\cite{steffens2020attribution}. The CKC provided a structured methodology for analyzing and responding to cyber threats. Concurrently, the MITRE ATT\&CK framework~\cite{mitreattack} emerged, offering a comprehensive knowledgebase of TTPs used by threat actors.
Further, the attribution framework such as Diamond Model~\cite{caltagirone2013diamond} and MICTIC framework~\cite{steffens2020attribution} presents a structured way to categorize attribution artifacts on a different basis and employ expert reasoning to compare with historical databases. These frameworks enabled a more granular analysis of cyber incidents and facilitated the identification of patterns and signatures associated with specific threat actors. The progression of these methodologies marked a significant leap in the ability to not only detect but also attribute cyberattacks with greater precision~\cite{mei2022review}.

Traditional methods rely heavily on the expertise of cybersecurity analysts and investigators. Security analysts meticulously examine digital forensic evidence, such as malware signatures, IP addresses, and domain names used during the attack period. This process often involved correlating technical indicators with known threat actor profiles, analyzing linguistic and behavioral patterns, and sometimes leveraging intelligence from law enforcement and intelligence agencies~\cite{goel2021attribution,steffens2020attribution,usdj2014us,pahi2019cyber,detlefsen2015cyber}. 
The traditional methods provide deep and contextual attribution information but face several limitations.
The primary limitation of framework-based attribution lies in its inherent limitations regarding speed and scalability. In an era of rapidly increasing cyber-attacks, the time-consuming nature of manual analysis becomes a significant bottleneck. The delay in attribution hampers the timely response to ongoing threats and affects the ability to deter future attacks.
Furthermore, the reliance on expert analysis is invaluable for understanding complex cyber espionage campaigns or sophisticated state-sponsored attacks, but introduces subjectivity, potential bias, and scalability issues in the attribution process. 

\subsection{Automated APT attribution}
\label{subsec:automatedattribution}

Traditional attribution's complexity and resource demands have led to the development and adoption of automated APT attribution methods. Automated attribution leverages advanced technologies, including ML, AI, and big data analytics, to enhance the speed and accuracy of identifying threat actors. By automating the analysis of vast amounts of data from diverse sources, such as network logs, threat intelligence feeds, and malware repositories, these systems can detect patterns and correlations that human analysts might miss. The benefits of automated APT attribution are manifold:
\begin{enumerate}[label=(\alph*)]
\item It significantly reduces the time required to identify and attribute attacks, enabling faster response and mitigation efforts.
\item Automated systems can continuously learn and adapt, improving their performance over time as they process more data and encounter new threat patterns. 
\item Automation can help overcome some of the limitations of manual methods, such as the ability to analyze large-scale and complex datasets that would be impractical for human analysts to handle. 
\item Automated methods reduce the likelihood of human error and bias in analysis, ensuring a more consistent and objective approach to attribution.
\end{enumerate}
As automated methods continue to advance, these methods are becoming indispensable for organizations aiming to identify threat actors and mitigate potential damage efficiently and quickly. Overall, the automated APT attribution represents a critical advancement in cybersecurity and offers a scalable and efficient approach to understanding and combating sophisticated cyber threats. In this paper, we used the word attribution and automated attribution interchangeably. 
\\
\\

\section{Attribution Case Study}
\label{sec:casestudy}
To demonstrate the APT attack investigation and artifacts aid in attribution, we discuss a high-profile target campaign performed in $2020$ employing supply chain-attack. This case study discusses the steps involved in tracing the attack, the techniques used by the threat actors, and the artifacts collected to accurately attribute the attack. By dissecting this high-profile incident, we aim to provide a comprehensive understanding of the methodologies and key artifacts which led to attribute this campaign.

\subsection{SolarWinds cyberattack}
\label{subsec:solarwind}
\begin{figure}
    \centering
    \includegraphics[width=\columnwidth]{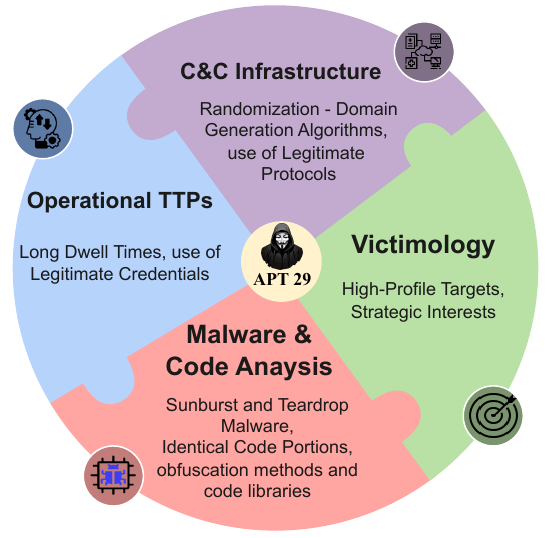}
    \caption{Solarwinds case study: attribution artifacts}
    \label{fig:solarwindcase}
\end{figure}

In December 2020, the leading cybersecurity firm FireEye (now Mandiant, a Google cloud company) discovered a widespread cyber espionage campaign.
This attack targeted the software supply chain of SolarWinds\footnote{A major IT management company} by compromising its Orion software to distribute malware to several high-profile organizations, including U.S. government agencies and private sector companies~\cite{MITRE_Solarwind}. 
Several leading security firms collaborated to share intelligence and uncover the responsible nation-states for such a large-scale cyber attack~\cite{krebsonsecurity2020}. 
The artifacts used by them to attribute the attack are shown in Fig~\ref{fig:solarwindcase} and described below.
\begin{enumerate} [label=(\alph*)]
    \item \textbf{Malware \& code analysis:} The key artifact in the this attack was the Sunburst malware embedded within the Orion software updates~\cite{sunburst2020}. After installation, it deploys secondary payloads like Teardrop malware, which enabled lateral movement within networks.
    \item \textbf{C\&C infrastructure:} The attackers employed a variety of domains for C\&C purposes such as advanced evasion techniques like randomization using DGA (Domain Generation Algorithm) and encryption, making detection and analysis challenging~\cite{solarwind2020analyzing}.
    \item \textbf{Operational TTPs:} The attackers' operational tactics also played a vital role in the attribution process. In the Solarwinds attack, the attackers demonstrated these advanced tactics by maintaining a low profile within the networks for several months, using legitimate credentials, and exfiltrating data without triggering alarms~\cite{solarwind2020tactics,MITRE_G0016}. 
    \item \textbf{Victimology:} The selection of targets provided additional context for attribution. The attackers focused on high-profile entities, including U.S. government agencies, critical infrastructure, and major corporations~\cite{MITRE_Solarwind}. 
\end{enumerate}

The SolarWinds campaign showcases the complexity and sophistication of advanced cyber espionage campaigns. By examining artifacts such as malware analysis, C\&C infrastructure, code overlaps, operational tactics, and victimology, cybersecurity experts attributed the campaign to APT29 (also known as Cozy Bear or Nobelium)~\cite{solarwind2020highly,MITRE_Solarwind}.
The reverse engineering revealed sophisticated coding, obfuscation techniques, and code similarities between the SolarWinds malware and previous APT29 campaigns~\cite{solarwind2020analyzing,sunburst2020}, providing key evidence for attribution. The use of DGA, randomization techniques, and the long dwell time were consistent with APT29's past behavior~\cite{solarwind2020highly}. The targeted intelligence-gathering aligned with APT29's strategic objectives, further supporting the attribution of the SolarWinds espionage campaign to the Russia-backed group APT29\cite{solarwind2020suspected,solarwind2020cisaattack}. The convergence of these multiple evidence provided a robust foundation for attribution. This analysis highlights the importance of comprehensive and multi-layered artifact analysis in identifying and understanding sophisticated threat actors.

\section{Systemization of Attribution Artifacts}
\label{sec:systemizationattributionartifacts}

Attribution is a multifaceted approach and done based on multiple clues. We call these clues as attribution artifacts.
We define \textbf{attribution artifacts} as \emph{any critical pieces of evidence or indicators or information which aid in attributing the cyber threats, i.e., linking cyber threats to specific threat actors or origins or sources}. During Russia's invasion of Ukraine, the AcidRain wiper malware was employed in the Viasat attack~\cite{cyber2022viasat,antony2022attribution}, and further analysis uncovered code overlaps between AcidRain wiper and the SolarWinds malware~\cite{solarwind2020highly}.

Artifacts of threat attribution hold significant importance because they serve as the foundational elements that identify and link cyber threats to specific actor or group. These artifacts, including malware signatures and attack behavior patterns, are crucial in linking an incident to its perpetrators. By examining these pieces of evidence, analysts understand the source and intent of an attack, which can directly influence how an organization or government responds to and mitigates potential threats.

\subsection{Attribution artifacts taxonomy}
\label{subsec:taxonomyartifacts}


Artifacts play a crucial role in the attribution process. It is essential to understand the various artifact types that can help security analysts to link or profile the threat actor. Such understanding enhances defenders' ability to trace attack origins and identify the techniques used. In this section, we present the taxonomy of artifacts as shown in Fig.~\ref{fig:ArtifactTaxonomy}.
Our taxonomy of attribution artifacts categorizes the diverse information that can be utilized in the cyber threat attribution process. 
Based on source of the attribution artifacts, we categorize them in two types: 1) Attack artifacts 2) Non-attack artifacts. 
The details about these artifact categories and further classification are discussed in this section.

\subsubsection{\textbf{Attack artifacts}} 
\begin{figure*}[!ht]
    \centering
    \includegraphics[width=23cm, height=16cm, angle=90]{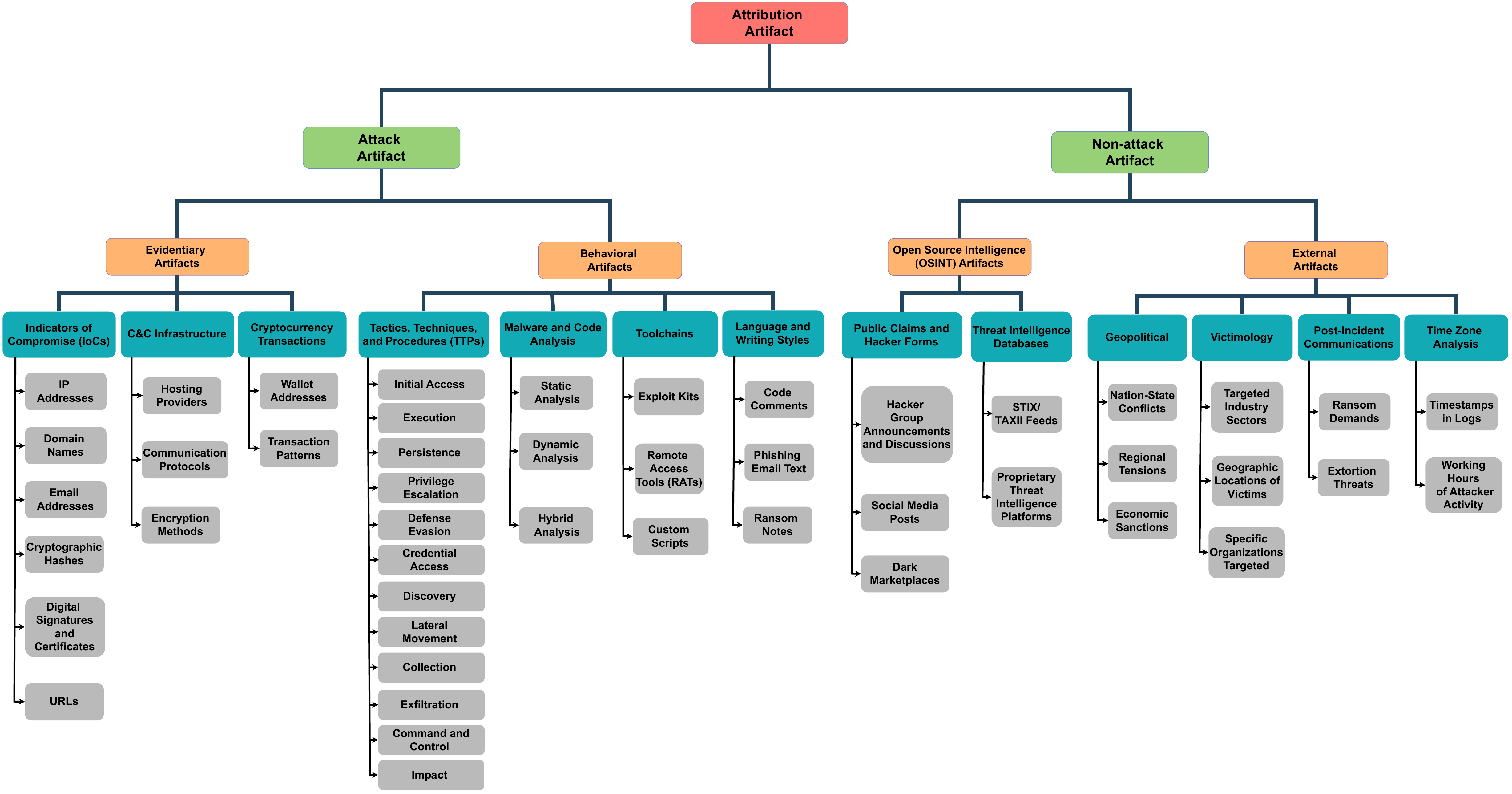}
        \caption{Taxonomy of attribution artifacts}
        \label{fig:ArtifactTaxonomy}
    \end{figure*}

It consists the artifacts which analysts captured from compromised infrastructure such as malware, IoCs, and toolchains etc. Based on nature of the artifacts, we categorize attack artifacts into two category: 1) Evidentiary artifacts and 2) Behavioral artifacts. The details about these artifact categories and further classification are discussed below.
\begin{enumerate}[label=\alph*)]
    \item \textbf{Evidentiary artifacts:} The tangible data or information providing direct evidence of a threat actor are evidentiary artifacts. These artifacts are crucial for identifying and analyzing specific attack elements as they can be directly observed and recorded. This includes direct digital evidence, such as IP addresses, cryptographic hashes, and email addresses, obtained from cyber incident investigations. Evidentiary artifacts offer concrete evidence that can link cyber activities to specific devices, networks, or individuals. We further divide evidentiary artifact into three types as explained below:
    \begin{enumerate}[label=\roman*)]
        \item \textit{Indicators of Compromise (IoCs):} Specific pieces of evidence indicating a system has been compromised are IoCs. A direct link to specific threat actors or campaigns can be made if same IoCs are captured in two attacks. Concrete examples of evidentiary artifacts may include IP addresses, URLs, email addresses, cryptographic hashes, domain names, digital signatures, and certificates.
        \item \textit{C\&C infrastructure:} It includes the servers and domains that control and manage deployed malwares. Unique characteristics of C\&C infrastructure can reveal the operational patterns of attackers, often tied to specific groups. This information details the resources attackers use to maintain communication with compromised systems—for example, hosting providers, communication protocols, and encryption methods.
        \item \textit{Cryptocurrency transactions:}  Transactions involving cryptocurrencies are often linked to ransom payments or the purchase of illicit services on the dark web. Blockchain analysis can sometimes trace these transactions to specific wallets and aid in connecting them to known groups or activities~\cite{ahmadi2019federated,korver2019attribution,cherniei2021criminal}. For example,  analyzing wallet addresses and transaction patterns links transactions to known wallets.
    \end{enumerate}

    \item \textbf{Behavioral artifacts:} The methods attackers use, including their operational patterns and decision-making processes, are known as behavioral artifacts. Understanding these behaviors can reveal the attackers' level of sophistication, potential motives, and operational goals. It is crucial to identify the likely perpetrators behind cyber attacks. We further divide behavioral artifact into four types as explained below:
     \begin{enumerate}[label=\roman*)]
        \item \textit{Tactics Techniques and Procedures (TTPs):} It describes how threat actors operate and offers valuable insights into their behavior and capabilities~\cite{daszczyszak2019ttp}. Specific TTPs can act as signatures for certain groups, revealing their unique modus operandi such as using phishing to perform initial access or employing PowerShell for execution. The TTPs can be extracted from historical threat reports~\cite{rani2023ttphunter,rani2024ttpxhunter,husari2017ttpdrill} or threat intelligence feeds to create the dataset. The TTP extraction methods help attackers to achieve their objectives and provide a detailed picture of their operational approach. The MITRE ATT\&CK framework includes $12$ different type of tactics as shown in Fig~\ref{fig:ArtifactTaxonomy}. 
        \item \textit{Malware \& code analysis:} The artifacts related to malware analysis and the source code of malware are crucial for understanding its functionality, propagation methods, and origins. Unique code snippets, code reuse, or targeted vulnerabilities can link malware to known threat actors~\cite{marquis2015big}. For example, unique malware functions and reusing specific code blocks seen in other known malware can provide valuable insights. The capabilities of the malwares can also be extracted and correlate them to identify similar ones~\cite{shenderovitz2024bon,saha2023malxcap,alrawi2021forecasting}. The process of dissecting malware helps identify its behavior and origin. It involves performing static, dynamic, and hybrid analysis of malware samples.
        \item \textit{Toolchains:} It includes the tools utilized by attackers throughout the lifecycle of an attack. The preference for certain tools or specific combinations can indicate particular threat actors or communities~\cite{gray2024identifying}. The custom-built hacking tools and well-known exploitation frameworks can provide clues about the attackers' identities and affiliations. These tools and software are essential for attackers to conduct the operations such as exploit kits, remote access tools, and custom scripts.
        \item \textit{Language and writing styles:} Linguistic analysis can indicate the native language of the attackers and potentially their country of origin~\cite{steffens2020attribution}. For example, using specific dialects or idioms in phishing emails and error messages in malware can provide valuable insights. These linguistic characteristics help identify and profile the attackers. 
        It includes code comments, phishing email text, and ransom notes.
    \end{enumerate}
\end{enumerate}

\subsubsection{\textbf{Non-attack artifacts}}
It consists the artifacts which analysts captured from outside of the compromised infrastructure such as threat intelligence feeds, public claims, and geopolitical tension etc. Based on nature of the artifacts, we categorize non-attack artifacts into two category: 1) Open Source INTelligence (OSINT) artifacts and 2) External artifacts. The details about these artifact categories and further classification are discussed below.
\begin{enumerate}[label=\alph*)]
    \item \textbf{Open Source INTelligence (OSINT) artifacts:} It includes the artifacts derived from publicly available sources, such as social media, threat reports, technical databases of historical attacks, and hacker forums. These artifacts can provide contextual and corroborative information to support attribution by linking public disclosures or activities to specific cyber incidents or threat actors~\cite{ren2022cskg4apt}. OSINT artifacts offer valuable insights into the threat actor’s activities, intentions, and associations.
    We further divide this artifact into two types, explained below:
    \begin{enumerate}[label=\roman*)]
        \item \textit{Public claims and hacker forums:} The online platforms, including hacker forums, dark websites, and social media channels, are crucial sources of information where cybercriminals discuss tactics, share tools, and claim responsibility for cyber incidents. Although these claims need verification, they can provide links to the perpetrators. For instance, threat group blogs on dark websites, telegram channels, and social media posts by groups claiming an attack are crucial sources of information~\cite{anonymous_sudan,telegram_investigation}. Information or tools shared on public platforms can often be directly linked to specific cyber incidents or groups, such as a thread in a dark web forum where an actor sells access credentials of compromised systems. 
        \item \textit{Threat intelligence databases:} The databases aggregate and analyze IoCs, attack patterns, and threat actor profiles. By correlating data from historical threat databases with indicators captured during an incident, analysts can establish links and identify related known threat actors~\cite{qiang2017framework,zhou2022cti}. These repositories of threat data are invaluable for identifying and understanding cyber threats. For example, STIX/TAXII feeds and proprietary threat intelligence platforms.
    \end{enumerate}

    \item \textbf{External artifacts:} It includes information about the broader context in which cyber threats occur such as geopolitical events and international cyber policies. These artifacts provide insights into the external motivations and constraints affecting cyber threat actors. By analyzing them, we can potentially identify why an attack was carried out and who might be responsible based on alignment with geopolitical motives. We categorize these artifacts into four types:
    \begin{enumerate}[label=\roman*)]
        \item \textit{Geopolitical:} Information related to international relations, conflicts, and political events that may motivate or be exploited by cyber threat actors is crucial. We can identify possible state sponsorship or politically motivated actors by correlating cyber attacks with geopolitical events~\cite{steffens2020attribution,egloff2023publicly}. A rise in cyber espionage targeting government agencies during periods of heightened diplomatic tensions between two countries can indicate geopolitical motives. 
        We witnessed such campaigns during recent Russia-Ukraine conflicts~\cite{Wiki_RussoUkrainianCyberwarfare,CSIS_CyberOperations}.
        \item \textit{Victimology:} The study of victims targeted by cyber-attacks helps to identify patterns that can reveal the attackers’ motives~\cite{hawdon2021cybercrime,crsreport2021}. Analyzing victim selection provides insights into the attackers’ objectives and potential identities, allowing us to link them to specific groups or goals. This has been thoroughly discussed in CAM (Cyber Attribution Model)~\cite{pahi2019cyber}.
        Key areas to examine include targeted industry sectors, the geographic locations of victims, and specific organizations that were attacked.
           \item \textit{Post-incident communication:} Communication after an incident can provide valuable insights into the attackers’ motives or negotiation strategies~\cite{schnirch2024exploring}. For instance, ransom notes, negotiation attempts, or taunting messages can reveal important details about the attackers’ intentions and tactics. These communications often occur directly between the attacker and the victim, offering a glimpse into the dynamics of the attack. Examples include ransom demands and extortion threats.
         \item \textit{Time zone analysis:} Analyzing the time zones in which attackers operate can provide clues about their origins and working patterns~\cite{steffens2020attribution}. By examining timestamps in logs and the hours during which attacker activity is most frequent, we can infer the likely geographic locations of the threat actors. This information helps narrow down potential suspects and understand the operational habits of the attackers. Examples include evaluating timestamps in system logs and identifying the working hours of attacker activity.
    \end{enumerate}
\end{enumerate}

\subsection{Comparison of automated attribution artifacts}
\label{subsec:comparisonartifacts}

\begin{table*}[!h]
    \centering
    \caption{Comparison of attribution artifacts}
    \label{tab:comparisonartifacts}
    \begin{tabular}{>{\raggedright}p{1cm}@{\hspace{0.6cm}}p{1cm}@{\hspace{0.8cm}}p{5cm}ccccc}
\hline
\textbf{Source based} & \textbf{Nature based} & \textbf{Artifact Name} & \textbf{Relevance} & \textbf{Integrity} & \textbf{Credibility} & \textbf{Timeliness} & \textbf{Accessibility}\\
\hline
\multirow{7}{1cm}{Attack Artifacts} & \multirow{3}{1cm}{Evidentiary Artifacts} & Indicator of Compromise (IoCs) & \high & \low & \high & \high & \high \\
& & C\&C Infrastructure & \high & \low & \high & \medium & \medium \\
& & Cryptocurrency Transaction & \low & \low & \high & \low & \low \\
[7pt]
& \multirow{4}{1cm}{Behavioral Artifacts} & Tactics, Techniques \& Procedures (TTPs) & \high & \high & \high & \medium & \medium \\
& & Malware \& Code Analysis & \high & \high & \high & \medium & \low \\
& & Toolchains & \medium & \low & \high & \medium & \medium \\
& & Language \& Writing Style & \low & \medium & \high & \low & \medium \\
\hline
\multirow{6}{1cm}{Non-attack Artifacts} & \multirow{2}{1cm}{OSINT Artifacts} & Public Claims and Hacker Forums & \medium & \low & \low & \medium & \high \\
& & Threat Intelligence Databases & \medium & \medium & \medium & \low & \high \\
[7pt]
& \multirow{4}{1cm}{External Artifacts} & Geopolitical & \medium & \medium & \medium & \low & \medium \\
& & Victimology & \high & \high & \high & \medium & \medium \\
& & Post-Incident Communication & \low & \low & \high & \low & \medium \\
& & Time Zone Analysis & \medium & \low & \high & \high & \high \\
[3pt]
\hline
\end{tabular}

\end{table*}
For accurate threat attribution, it is crucial to analyze and compare the effectiveness of various artifacts~\cite{hwang2023exploratory,skopik2020under}. Different artifacts, such as malware signatures, network traffic logs, and OSINT data, provide unique insights into an attack. Security analysts can discern patterns and link behaviors to specific threat actors by systematically evaluating these artifacts. 
Therefore, we compare the classified artifacts (as shown in Fig.~\ref{fig:ArtifactTaxonomy}) based on the different metrics: Relevance, Integrity, Credibility, Timeliness, and Accessibility. We compare artifacts based on these metrics with respect to three scale points: Low (\low), Medium (\medium), and High (\high). These scale points represent confidence in corresponding measures, i.e., low means minimal confidence, medium represents moderate confidence, and high means more confidence. Table~\ref{tab:comparisonartifacts} shows the comparison of all classified artifacts.

Relevance measures how directly an artifact contributes to the identification of a specific threat actor. High relevance artifacts, such as IoCs, TTPs, C\&C infrastructure, and malware \& code analysis, provide strong and often unique identifiers that can clearly tie activities to specific threat actors, making them crucial in the attribution process. Medium relevance artifacts, like toolchains, public claims, threat intelligence database, geopolitical, and time zone analysis, offer valuable context but may require additional corroboration to be definitive, reflecting their role in supporting rather than leading attribution efforts. Low relevance artifacts, such as cryptocurrency transactions, language and writing styles, and post incident communication, provide useful supplementary information but often lack the direct connection needed for strong attribution, making them less central to the process. This analysis aligns with the idea that artifacts vary in their directness and specificity when attributing cyber threats.

Integrity assesses the resistance of an artifact to manipulation or deception. High integrity artifacts like TTPs, victimology, and malware \& code analysis are reliable because they are challenging for attackers to fake or alter without compromising their operational capabilities, making them robust indicators for attribution. Medium integrity artifacts, such as language \& writing style, threat intelligence databases and geopolitical analysis, provide useful information but can be subject to interpretation, bias, or imitation, requiring careful validation. Low integrity artifacts, like IoCs, C\&C infrastructure, toolchain and public claims, are more easily manipulated or spoofed, reducing their reliability in making confident attributions. This analysis underscores the importance of using artifacts with high integrity to ensure accurate and reliable cyber threat attribution.

Credibility refers to the reliability of the source from which an artifact is obtained. It measures how likely it is that the information provided by the source is accurate and unbiased. High credible artifacts, such as those providing TTPs, C\&C infrastructure data, TTPs, toolchains,  and malware analysis, are typically derived from detailed investigations by reputable cybersecurity professionals and organizations, ensuring the reliability and accuracy of the information. Medium credible artifacts, such as geopolitical analysis, and threat intelligence databases, offer useful insights but may come from a mix of reliable and less verified sources, requiring cross-verification to ensure accuracy. Low credible artifacts, such as public claims and hacker forums, are often unverified and can be intentionally misleading, making them less trustworthy for confident attribution. This analysis emphasizes the need to prioritize artifacts from high-veracity sources to ensure accurate and credible cyber threat attribution.

Timeliness measures how quickly an artifact can be obtained and analyzed after an attack. High timeliness artifacts, such as IoCs, and time zone analysis, are often available almost immediately during or after an attack, allowing for rapid response and mitigation. Medium timeliness artifacts, like C\&C infrastructure, TTPs, victimology, public claims, and malware \& code analysis typically require some time for monitoring, analysis, and pattern recognition, which may delay immediate action but still offer relatively prompt insights. Low timeliness artifacts, such as cryptocurrency transactions, language \& writing styles, threat intelligence databases, post incident communications and geopolitical considerations, involve more extensive investigation and slower information gathering, making them less suited for real-time threat response. This analysis emphasizes the importance of prioritizing timely artifacts in situations where rapid attribution and response are critical.

Accessibility refers to how easily an artifact can be collected and analyzed during the attribution process. High accessibility artifacts like IoCs, public claims, threat intelligence databases and time zone analysis are readily available and require minimal effort to collect and analyze, making them easier for defenders to leverage in attribution. Medium accessibility artifacts, such as TTPs, C\&C infrastructure,  toolchains, language \& writing analysis, geopolitical, victimology, and post incident communications involve a moderate level of complexity and require some specialized knowledge and tools, reflecting the need for more effort to effectively utilize them. Low accessibility artifacts, such as malware \& code analysis and cryptocurrency transactions, demand advanced technical skills and resources, akin to the top of the Pyramid of Pain, where the effort to analyze and attribute becomes more challenging and resource-intensive. This analysis highlights how the accessibility of artifacts aligns with the effort required to disrupt or attribute adversarial activities.

This comparative analysis enhances the reliability of attribution, enabling more precise identification of adversaries. A glimpse of the reasoning behind this analysis is also provided in Table~\ref{tab:artifactcomparisonexplaination} in the Appendix. Moreover, understanding the strengths and limitations identified during comparison of each artifact type ensures a comprehensive approach, reduces the likelihood of false attributions, and improves overall threat intelligence.

\section{Systemization of Automated Attribution Methods}
\label{sec:systemizationattributionmethods}

In this section, we systemize automated attribution methods, detailing each aspect comprehensively. First, we outline the framework followed for automated attribution. Next, we discuss publicly available datasets that facilitate the analysis and development of attribution methods. We then delve into the current literature on automated attribution methodologies, providing a detailed categorization of datasets and methods. Finally, we discuss critical comments and comparisons of existing automated attribution methods, mentioning their strengths and limitations.

\subsection{Automated attribution framework}
\begin{figure}[!h]
    \centering
    \includegraphics[width=\columnwidth]{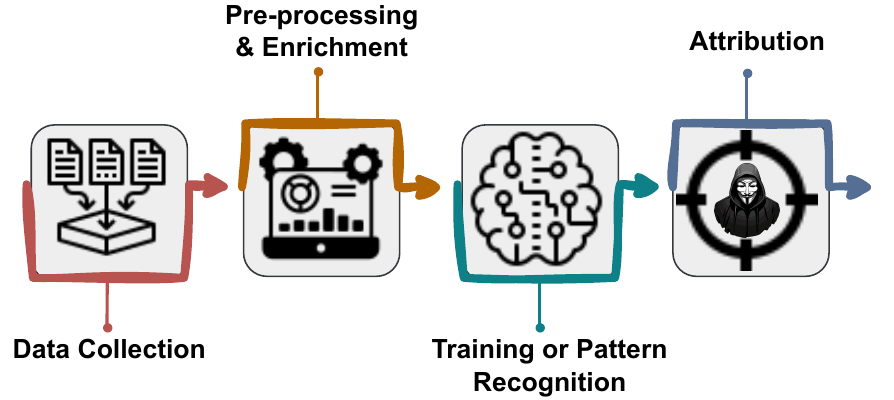}
    \caption{Automated APT attribution framework}
    \label{fig:attribtuionframework}
\end{figure}

APT attribution leverages advanced technology to automate the process of identifying the actors behind APT campaigns without or with minimal human intervention.
The process leverages AI, ML, and big data analytics to analyze the TTPs and other artifacts used in cyber attacks. The analysis includes comparison or correlation of given artifacts against extensive databases of known threat actors and their historical activities. The goal is to rapidly and accurately identify the perpetrators by understanding their motives, methodologies, and potentially their affiliation. 
The typical automated APT attribution process involves four key steps as shown in Fig~\ref{fig:attribtuionframework} and described below:
\begin{enumerate}[label=(\alph*)]
    \item \textbf{Data collection:} The first step includes gathering the vast amounts of data from various sources, including network traffic, logs, malware samples, and threat intelligence feeds. These data provide the raw material for analysis.
    \item \textbf{Preprocessing and enrichment:} The collected data is cleaned, normalized, and enriched to facilitate analysis. This step may involve extracting features from raw data such as malware samples, network and system logs, correlating related data points, and filtering out irrelevant information.
    \item \textbf{Training or pattern recognition:} Using ML algorithms, the system analyzes the preprocessed data to identify patterns and behaviors that match known APT groups. This analysis can include the examination of malware signatures, communication patterns, and attack methodologies.
    \item \textbf{Attribution:} Once analysis or training is performed, the system attributes the attack to specific APT groups by matching identified patterns with known threat actors. This step relies on a comprehensive database of threat actor profiles, including their TTPs, preferred malware, targets, and historical attacks.
\end{enumerate}
The automated attribution process begins with extensively curating APT data from both public and private repositories. Next, preprocessing and feature engineering are conducted. The engineered features are then utilized to learn the attribution aspects, culminating in the final attribution process.

\subsection{APT attribution dataset and knowledge base classification}
\label{subsec:aptdatasets}

We curate the publicly available datasets and knowledge bases, which can be used to develop APT attribution methodology.  In our exploration, we classify public data sources based on the type of data they consist. Therefore, we classify the data repositories into four types: Malware dataset, Threat report dataset, Attack pattern dataset, and Heterogeneous dataset. The distribution is shown in Fig.~\ref{fig:datasetclassification}. We find a total of twelve public sources that facilitate data collection specific to APT attacks, which we discuss in detail below:
\begin{figure}[!h]
    \centering
    \includegraphics[width=6cm, height=6cm]{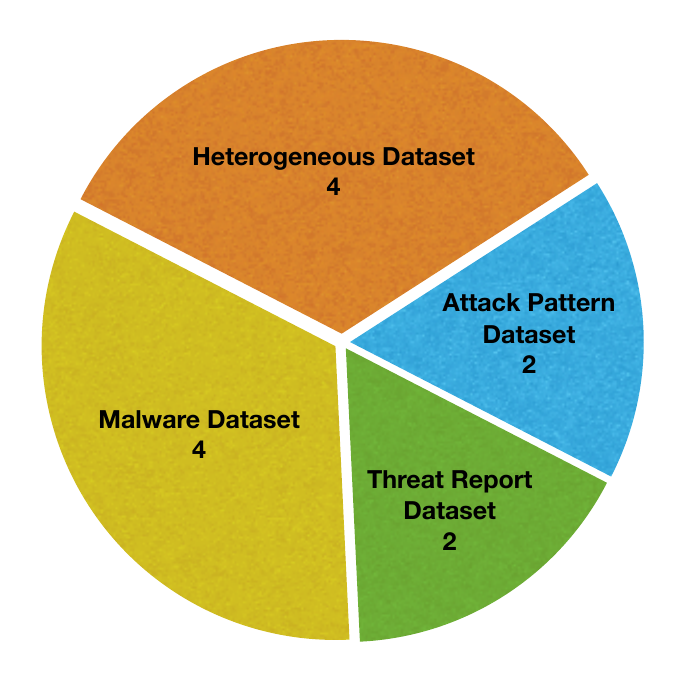}
    \caption{Dataset and knowledge base classification of APT attribution}
    \label{fig:datasetclassification}
\end{figure}
\\
\subsubsection{Malware dataset}
\label{subsubsec:malwarebaseddata}

Researchers extensively utilize malware associated with APT groups from past attack campaigns to develop attribution methodologies, such as~\cite{xu2021apt,kida2022nation,rosenberg2017deepapt,wang2021explainable,li2021attribution,wang2022apt,haddadpajouh2020mvfcc,kim2021automatically,lee2022malware,mei2022hybrid}. To develop the methodology, several researchers have gathered malware samples from various repositories and labeled their corresponding APT groups using OSINT databases. We name such collection of APT malware along with their associated threat group dataset as malware dataset for attribution. We discuss all such public dataset collections in this section.

The authors in~\cite{boot2019applying,bootsupervised} compiled an diverse and balanced APT malware dataset using malware samples and their corresponding sandbox reports. They obtained details on state-sponsored malware, like sample hashes and corresponding threat group name, from threat intelligence reports published by security firms like Trellix~\cite{trellix}, F-Secure~\cite{fsecure}, and Kaspersky~\cite{kaspersky}.
They leveraged VirusTotal~\cite{VirusTotal} to download samples and prepare dataset of $3,594$ unique samples labeled to $12$ state-sponsored APT groups. The dataset is made publicly available~\cite{APTMalware} for further research.


The study of Gray et al.~\cite{gray2024identifying} introduces APTClass, a dataset containing $17,513$ labeled malware samples prepared for research in malware-based attribution.
APTClass includes ELF, Windows EXE, and DLL files, prepared by extracting text from PDF reports, YARA rules, and IoC files, performing hash searches for sample collection, and using metadata N-gram and keyword searches for APT label extraction.
The dataset can be enriched with OSINT indicators and is available at \cite{aptclass}.

Haddadpajouh et al.~\cite{haddadpajouh2020mvfcc} present a dataset of $1,200$ samples from five APT groups (APT1, APT3, APT28, APT33, and APT37) collected from various public repositories.
They leverage the cuckoo sandbox reports to prepare multiple views for each sample, focusing on opcode, bytecode, system call, and header information. This approach captures a multi-dimensional representation of the malware, allowing for an in-depth analysis across several facets.
This dataset is made publicly available~\cite{cyberscienceAPTmalware} for the benefit of the community. 
\\
\subsubsection{Threat report dataset}
\label{subsubsec:threatreportdata}
Research has also been done using past incident threat reports to perform threat attribution, such as~\cite{perry2019no,naveen2020deep}. We explore collection of threat reports available publicly and categorize them in the threat report dataset type. The objective of such dataset is to collect attack information, including attack flow, used tools and techniques, and modus opernadi of threat actors to let model profile threat actors behaviour and aids in threat attribution. 

Our exploration find a frequently used threat report collection of APTs named APTNotes~\cite{aptnotes}. It constitutes an organized repository that aggregates a diverse range of public threat reports, whitepapers, and articles dedicated to APT campaigns range from $2008$ to $2023$.
The repository provides direct links and also maintains archived versions of these resources. 

We find a repository APT Cyber Criminal Campaign Collection~\cite{aptcybercriminal}, a vital resource for cybersecurity research. It offers an extensive archive of APT group and cybercriminal campaign information, centralizing documented incidents for a comprehensive understanding of cyber threats and adversary tactics from 2008 to present. This repository aids researchers and cybersecurity professionals by providing open access to detailed analyses of cybercriminal activities to support the development of defense strategies. 
This repository is up-to-date and continuously being updated by esteemed contributors.
\\
\subsubsection{Attack pattern dataset}
\label{subsubsec:attackpatternbaseddata}
Researchers have also leveraged the modus operandi of threat actors in terms of TTPs to profile the threat actors and perform attribution such as~\cite{noor2019machine,edie2023extending,shin2021art}. The frequent list of TTPs represents the common patterns a specific threat actor exhibits. We classify such APT group’s TTP dataset as the attack pattern dataset.

Edie et al.~\cite{edie_mckee_attribution_2023} provide a comprehensive list of extended threat playbooks, i.e., a set of TTPs that an APT group regularly uses. 
They use association rule mining with the aid of the apriori algorithm to learn the statistical relationships between TTPs. For example, if T1027 and T1059 are observed, then T1105 is hypothesized to exist in the same attack campaign. The hypothesized TTP helps analysts hunt for unobserved TTPs and draw analytical conclusions around attack scenarios. This dataset contains $1,316$ observations and publicly available~\cite{extendingaptattributiondata} for community benefits.

MITRE ATT\&CK framework~\cite{mitreattack} is a knowledge base which lists a set of TTPs used by threat groups in past attributed attack campaigns. We can map the TTPs associated with a specific threat group through ATT\&CK framework~\cite{attacknavigator}.
\\
\subsubsection{Heterogeneous dataset}
\label{subsubsec:heterogenousbaseddata}

This dataset is the collection of numerous data types and their corresponding APT groups, which has been leveraged to propose threat attribution process~\cite{noor2023machine,sachidananda2023apter,xiao2024apt,xu2022hghan,wu2020grouptracer}. 

Noor et al.~\cite{noor2023machine} developed a dataset derived from publicly accessible CTI documents from leading IT and cybersecurity firms such as Mandiant~\cite{mandiant}, Unit42~\cite{unit42}, and Crowdstrike~\cite{crowdstrike}. This dataset captures low-level IoCs like malware hashes and IP addresses from CTI reports on various cyber threat actors. Additionally, it incorporates high-level IoCs from the MITRE ATT\&CK framework, which details adversaries’ TTPs. The creation process involved manually extracting low-level IoCs and synthesizing high-level IoC data to include missed or inaccurately attributed IoCs. This synthesis aimed at a more accurate cyber threat attribution by allowing multiple instances per cyber threat actor. The resulting dataset, comprising low-level and high-level IoCs, seeks to improve cyber threat attribution performance and is available for further study~\cite{noorattackpatterndata}.

An Excel sheet focused on APT groups and their operations~\cite{aptgroupoperations} is also a knowledge base and critical asset for cybersecurity analysts and researchers. It encompasses a wealth of data, including identifying APT groups, their geographical or organizational origins, details on TTPs employed, associated malware tools, descriptions of known campaigns, and IoCs. This organized compilation facilitates a comprehensive understanding of the threat landscape posed by these sophisticated actors, detailing their methodologies, and the specific vulnerabilities they exploit. By aggregating and categorizing this information, the document aids in identifying, analyzing, and mitigating threats from various APT groups, enhancing the overall cybersecurity posture against such entities.

MITRE's knowledge base curates the extensive details related to activities performed by threat groups~\cite{mitregroup}.
It endeavors to map APT groups to their TTPs~\cite{mitreattack}, as well as the softwares~\cite{mitresoftware} attackers utilize and the campaigns~\cite{mitrecampaign} they conduct, and various naming aliases which acknowledge that different organizations may track the same entities under different names. This approach aids in understanding the overlaps and distinctions among groups, enhancing analyst awareness and encouraging further investigation. Software in MITRE framework encompasses a variety of applications, ranging from custom malware to commonly used operating system utilities, categorized into malware and tools based on their intended use by adversaries. This categorization helps understand the specific techniques that different software instances can employ, mapped to the groups known to use them. Campaigns are defined as structured series of intrusion activities with specific objectives, targets, and duration. MITRE records these activities, providing unique identifiers for unnamed operations or using publicly reported names, which aims to attribute them to known groups and their used software where possible. MITRE facilitates a comprehensive view of threat actor’s operations over time and offers insights into the evolution of their tactics and methodologies.

Laurenza and Lazzeretti introduce “dAPTaset” ~\cite{laurenza2020daptaset}, a structured database for APT information. The database structure includes APTs, keywords, threat reports, samples, network information, CVEs (Common Vulnerabilities and Exposures), attack techniques, and more. The data collection process involves multiple sources such as MITRE ATT\&CK~\cite{mitreattack}, APT groups and operations spreadsheets~\cite{aptgroupoperations}, APTNotes~\cite{aptnotes}, and MISP galaxy cluster~\cite{MISP_Galaxy}, from which APT names, aliases, victims, operations, and external references are extracted. They parsed the public threat reports to gather IoCs, file hashes, IP addresses, network URLs, emails, CVEs, and keywords related to APTs. The dataset is further enriched by cross-referencing with cybersecurity websites like VirusTotal~\cite{VirusTotal}, MalwareBazaar~\cite{malwarebazaar}, VirusShare~\cite{VirusShare}, Malshare~\cite{MalShare}, and Malpedia~\cite{Malpedia} to validate and supplement information. This dataset uses $821$ reports of $88$ different threat groups. In addition, it also includes a total of $21,841$ binary hashes, $2,620$ network-related IoCs, $8,927$ unique files, $169$ CVEs, and $175$ different techniques related to APT activities. The dataset is publicly available~\cite{dAPTaset} for the benefit of the community.
\\
\\
The discussed datasets are crucial for gathering threat actors’ characteristics and developing robust attribution methods. Using these datasets, researchers can cross-reference and validate threat intelligence, identify patterns, and attribute cyber incidents to specific threat actors with higher precision. Integrating these diverse data sources is essential for developing robust threat attribution methodologies. A comparison of discussed datasets is present in Table~\ref{tab:Dataset}.    

\begin{table*}[!ht]
        \centering
        \caption{Comparison of the datasets}
        \label{tab:Dataset}
        \begin{threeparttable}
\begin{tabularx}{\textwidth}{XXXXXX}
\hline
\textbf{Dataset} & \textbf{Type} & \textbf{No. of sample} & \textbf{No. of APT group} & \textbf{No. of groups more than 25+ samples} & \textbf{No. of groups more than 50+ samples} 
    \\
    \hline 
     APTMalware~\cite{boot2019applying} & Malware & $4,449$ & $12$ & $12$ & $11$ 
     \\
     APTClass~\cite{gray2024identifying}\tnote{*} & Malware & $17,513$ & $275$ & $85$ & $53$ 
     \\
     APTNote~\cite{aptnotes} & Threat Report\tnote{a} & $687$\tnote{b} & $-$ & $-$ & $-$ 
     \\
    dAPTaset~\cite{laurenza2020daptaset} & Heterogeneous & $8,927$ & $88$ & $14$ & $10$ 
    \\
     APT Dataset~\cite{haddadpajouh2020mvfcc} & Malware & $1,200$ & $5$ & $5$ & $5$
     \\
     AttackAtttribution~\cite{perry2019no}& Threat Report & $238$ & $12$ & $2$ & $1$
     \\
     Edie et al.~\cite{edie_mckee_attribution_2023} & Attack pattern & $1,316$ & $147$ & $0$ & $0$
     \\
     APT Cyber Criminal Campaign Collection~\cite{aptcybercriminal} & Threat Report\tnote{a} & $1,563$\tnote{b} & $-$ & $-$ & $-$ 
     \\
     \hline
\end{tabularx}

        \begin{tablenotes}
        \item[a] This is the curated list of unlabeled threat reports details along with threat report in pdf format.
        \item[b] This dataset is continuously being updated. So, the number is reported based on current samples in the dataset.
        \item[*] Available upon request. The more details can be found at \url{https://s3lab.isg.rhul.ac.uk/project/aptclass/}
        \end{tablenotes}
        \end{threeparttable}
    \end{table*}


\subsection{Classification of automated attribution methods}
\label{subsec:classattributionmethod}

We classify the attribution methods in the literature based on dataset and algorithm types. 
Based on the type of algorithm, we classify the present attribution methods into classification-based, clustering-based, and similarity-based. The classification-based methods are supervised ML methods trained on the labeled datasets, whereas clustering-based methods, i.e., unsupervised methods, are developed on the unlabeled datasets. Additionally, some work is based on matching the similarity between a given sample set with an established threat group database set rather than training any ML model.
Based on the type of dataset, we classify available attribution literature methods into four categories: Malware-based methods, Threat report-based methods, Attack pattern-based methods, and Heterogeneous-based methods. Malware-based methods leverage malware samples to perform attribution, Attack pattern-based methods leverage TTP to link threat actors, Threat report-based methods use threat intelligence reports' texts to perform attribution, and Heterogeneous-based methods consider diverse data types to establish the connection between given data and threat actor.  

Following the classification based on the dataset type, we comprehensively discuss the literature mentioned in Table~\ref{tab:classificationofmethod}. The details of the literature methods are explained below:

\subsubsection{Malware-based methods}
\label{subsubsec:malwaremethods}
To leverage malware for threat attribution, various researchers analyze malware’s several features and ML models to perform threat attribution~\cite{xu2021apt,kida2022nation,rosenberg2017deepapt,wang2021explainable,li2021attribution,wang2022apt,haddadpajouh2020mvfcc,kim2021automatically,lee2022malware,mei2022hybrid}. We review current malware-based threat attribution literature and discuss them in detail below.

Xu et al.~\cite{xu2021apt} present a combined feature set of TF-IDF and bi-gram methods prepared by API call sequences of APT samples. 
The combined features are huge in number, so they employ the Adaboost feature selection method to rank the features based on their importance and contribution to classification. They select the top-400 feature set as input and deploy several multiclass ML classifiers, i.e., DT, KNN, LightGBM, and XGBoost, to classify malware to their corresponding APT groups. Their result finds LightGBM as the best performer compared to all implemented classifiers.

Kida et al.~\cite{kida2022nation} present a novel way to leverage malware’s fuzzy hashes to attribute APT groups. They calculated five types of fuzzy hashes: TLSH, SSDEEP, SDHASH, IMPFUZZY, and LZJD for all malware samples and used computed hash as natural language input to the classifier. After hash calculation, the hashes are encoded using the NATO encoding method, and they use the padding method to keep the encoded length the same. Further, they use character level and n-gram vector embeddings to convert encoded hashes into feature vectors. The embedding vectors are forwarded to four different ML classifiers: RF, SVM, KNN, and NB. Their result demonstrates that RF with IMPFUZZY hash performs better than all implemented methods. 

\begin{table}[!h]
    \centering
    \caption{Classification of automated attribution methods}
    \label{tab:classificationofmethod}
    \begin{tabularx}{\columnwidth}{p{2cm}p{1.7cm}X}
    \hline
     \textbf{Base} & \textbf{Type} & \textbf{Paper} \\
     \hline
     \multirow{8}{*}{Algorithm Type} & Classification-based & ~\cite{rosenberg2017deepapt,perry2019no,noor2019machine,haddadpajouh2020mvfcc,naveen2020deep,xu2021apt,wang2021explainable,li2021attribution,kida2022nation,lee2022malware,mei2022hybrid,xu2022hghan,noor2023machine,sachidananda2023apter,xiao2024apt}  
     \\
     \\
       & Clustering-based & ~\cite{wu2020grouptracer,kim2021automatically,wang2022apt} 
      \\
      \\
      & Similarity-based  & ~\cite{shin2021art,edie2023extending} \\ 
      \hline
      \multirow{13}{*}{Dataset Type} & Malware-based & ~\cite{lee2022malware,wang2022apt,kida2022nation,li2021attribution,wang2021explainable,xu2021apt,haddadpajouh2020mvfcc,rosenberg2017deepapt} \\ \\
      & Threat report-based & ~\cite{xiao2024apt,wang2021explainable,naveen2020deep,perry2019no} \\ \\
      & Attack pattern-based & ~\cite{edie2023extending,shin2021art,kim2021automatically,wu2020grouptracer,noor2019machine}\\ \\
      & Heterogeneous data-based & ~\cite{sachidananda2023apter,noor2023machine,xu2022hghan,mei2022hybrid} \\
     \hline
\end{tabularx}

\end{table}

Rosenberg et al.~\cite{rosenberg2017deepapt} propose a method named DeepAPT to employ raw dynamic malware features and attribute APT groups linked with the given malware sample. They execute samples in a sandbox environment and collect behavior reports for each malware sample. DeepAPT considers words present in behavior reports as natural language features and evaluates the frequency of each unique word in all behavior reports. The authors selected the top $50,000$ frequent words as features and evaluated one-hot encoding to prepare the dataset. Further, they employ a 10-layer DNN to learn patterns from raw dynamic features for APT group classification. 

Wang et al.~\cite{wang2021explainable} introduce a novel approach for APT attribution by combining code and string features using paragraph vectors and bag-of-words vectors, respectively. They employ a RF classifier and DNN for classification by leveraging both types of features. 
Further, they add model interpretation techniques using Local Interpretable Model-agnostic Explanations (LIME), which offer insights into the classification process, enhancing trust and facilitating analysis for cybersecurity personnel.

Li et al.\cite{li2021attribution} propose an ML-based classification method for attributing APT malware to specific organizations in the IoT. 
Their method analyzes samples dynamically and extracts features from the dynamic analysis report. They extract the first $10k$ words and weigh each word using the TF-IDF method to transform features into vector form. Their method also includes filtration of stop-words before feature transformation. Out of all behavioral features, they employ the chi-square method to select features and use SMOTE (Synthetic Minority Oversampling Technique) to balance the dataset. Further, they leverage the RF classifier to perform multiclass classification between APT groups for a given malware sample.

Wang et al.~\cite{wang2022apt} introduce a novel method for automating the attribution of APTs by analyzing binary code. Their method identifies and classifies APT malicious codes by utilizing local features within binary functions and employing time series mining techniques. They extract $18$ different features of basic blocks from the CFG of the executable. They also filter the non-relevant functions to focus on those containing API calls indicative of malicious behavior. Next, they generate paths of interest based on API calls and train the PCA (Principal Component Analysis) to reduce the dimension from $18$ to $1$. They use HDBSCAN (Hierarchical Density-Based Spatial Clustering of Applications with Noise) to choose paths based on cluster classes and generate time series shapelets to represent key subsequences of paths. Next, they employ an RF classifier to attribute APT organizations to malware samples based on the distances between paths and path shapelets. 

Haddadpajouh et al.~\cite{haddadpajouh2020mvfcc} introduce a multi-view fuzzy consensus clustering model to attribute cyber threat payloads, specifically malware, to their respective actors.
Their study follows a multi-view approach (multiple pieces of evidence) rather than single-view (single piece of evidence) analyses. The multi-view is generated using four categories of features: Header, System call, Byte code, and Opcode. The authors calculate multi-view for each feature category using binary vector (1 if present 0 if absent), count vector, frequency of occurrence, TF-IDF vector, and eigenvector of CFGs. Further, they utilize fuzzy pattern trees and multi-modal fuzzy classifiers for inference. Their research demonstrates that a multi-view approach is significantly better than single-view analyses for cyber threat attribution.

Kim et al.~\cite{kim2021automatically} present a method for automatically attributing mobile APT groups by leveraging a vectorized ATT\&CK matrix and IoC pairing. It involves two primary steps: first, collecting and mathematically modeling TTPs from the MITRE ATT\&CK framework, followed by vectorization and storage for similarity measurement; second, enhancing attribution accuracy by comparing IoC pairs to counteract potential false flags by attackers. They use Joe Sandbox to extract ATT\&CK TTPs from malware. The experiments involve comparing the cosine similarity of the ATT\&CK matrix TTPs and the IoC pairings among the malware samples.

Lee and Cho~\cite{lee2022malware} propose a model for malware authorship attribution that leverages runtime modules to attribute linked APT groups. They perform frequency analysis to execute the samples to collect runtime-module names like Kernel32.dll and Advapi32.dll. They leverage one-hot encoding to convert features into numerical vectors and feed the vector to a set of ML classifiers for classification. They find the XGBoost algorithm as the best-performing model for attribution.

Mei et al.~\cite{mei2022hybrid} present a hybrid approach combining Particle Swarm Optimization (PSO) and Multiclass SVM (MSVM) for APT organizations by analyzing traces of APT attack tools in a sandbox environment. The approach optimizes MSVM parameters for accurate APT organization identification. They consider hybrid information as feature set, including file\_type, malware\_type, label, size, target\_object, etc. They use the TF-IDF method to prepare feature vectors and PSO method to initialize the parameters of MSVM. Further, they train the classification model to identify APT organization.
\\
\subsubsection{Threat report-based methods}
\label{subsubsec:threatreportmethods}
Some researchers leverage texts of past attack incidents threat reports to distinguish between campaign details of APT groups~\cite{perry2019no,naveen2020deep}. We reviewed these methods and discussed them in detail below.

Perry et al.~\cite{perry2019no} present NO-DOUBT, a novel ML-based method for attack attribution leveraging textual analysis of threat intelligence reports. They propose a new text representation algorithm, SMOBI (Smoothed Binary vector), designed to minimize feature engineering while capturing the contextual nuances of words through vector space representation derived from a labeled dataset and a vast corpus of security literature. They use report text as a feature, which is better for identifying similar reports than pinpointing threat actor behavior. SMOBI focuses on word frequency and similarity, ignoring semantics and context. Their approach also allows for identifying known and novel threat actors by classifying text into specific actor categories or signaling new, previously unseen actors.

Naveen et al.~\cite{naveen2020deep} introduce a deep learning architecture leveraging DNNs and domain-specific word embedding for attributing threat actors based on threat intelligence reports.
The novel vector representation method, SIMVER, utilizes semantic relationships within the text to improve attribution performance significantly. They tested their method against traditional methods using a dataset of threat reports~\cite{naveen2020deep}. Their method SIMVER demonstrates better performance in accurately identifying threat actors.


\subsubsection{Attack pattern-based methods}
\label{subsubsec:attackaptternmethods}

Various researchers have used threat actor's attack patterns in terms of TTPs to profile threat groups and perform attribution~\cite{noor2019machine,edie2023extending,shin2021art}. We discuss these methods in detail below.

Noor et al.~\cite{noor2019machine} propose an automated framework for cyber threat attribution by profiling threat actors based on their high-level attack patterns extracted from CTI reports. They develop a semantic search system based on the statistical distributional semantic relevance technique and map them according to ATT\&CK framework TTPs. Further, they prepare a feature correlation matrix for each threat report and employ ML classifiers to classify threat groups. They also observe the impact of features using the Information Gain (IG) feature selection method.

Edie et al.~\cite{edie2023extending} present a novel approach to cyber attack attribution using association rule mining and a weighted Jaccard similarity index.
Their approach involves learning associations between TTPs using MITRE ATT\&CK framework. These associations are represented as activity groups (AGs) using the apriori algorithm, aiding in identifying statistically related TTPs. The authors generate the standard and extended threat playbooks by incorporating confidence levels for known and hypothesized TTPs.
To identify potential adversaries, the algorithm is seeded with observed TTPs. Further, they use a weighted Jaccard similarity index to compare observed TTPs with threat playbooks and attribute attacks to likely adversaries.

Shin et al.~\cite{shin2021art} introduce Automated Reclassification for Threat Actors (ART) to automatically compare the TTPs from different APT groups by leveraging the MITRE ATT\&CK framework.
Their method involves crawling cyber threat reports and retrieving the ATT\&CK matrix of APT groups, preparing TTP-vectors, and calculating the cosine similarity to re-examine and classify the APT groups. 
They prepare $12$ different one-hot encoding vectors for each tactic in the ATT\&CK framework (ignores Pre-ATT\&CK tactics). Further, they calculate cosine similarity tactic-vector-wise between all groups and employ a weighting mechanism across the similarity score of all $12$ tactic vectors.
\\
\subsubsection{Heterogeneous data-based methods}
\label{subsubsec:heterogeneousmethods}
Along with the single type of data, i.e., malware, TTP, or threat report, several researchers have leveraged multiple types of data merged to perform attribution~\cite{noor2023machine,sachidananda2023apter,xiao2024apt,xu2022hghan,wu2020grouptracer}. They merge all relevant data about a threat actor and then use it as a dataset to develop a method for attribution. We discuss all of them in detail below.

Noor et al.~\cite{noor2023machine} leverage two types of high-level IoCs, i.e., TTPs and Software used by threat actors, to perform attribution. They leverage the list of TTPs present in ATT\&CK framework~\cite{mitreattack} and software provided by MITRE~\cite{mitresoftware} to prepare the dataset. They extract these high-level IoCs from threat reports and prepare feature vectors using the one-hot encoding method. Further, they employ several ML and DL classifiers to perform attribution and select the best-performing classifier as the final attribution model. They also compare attribution performance based on high-level and low-level IoCs and find that high-level IoCs are more trustworthy than low-level IoCs for threat attribution.

Sachidananda et al.~\cite{sachidananda2023apter} present methods named APTer, an innovative framework designed to tackle the complexity of APTs by correlating, predicting, and attributing stages of APT attacks, as well as mapping CVEs to the MITRE ATT\&CK framework. 
Their multifaceted approach includes eliminating redundant alerts, clustering related alerts to identify attack scenarios, and using ML models to predict future attack stages and attribute attacks to known APT groups based on their TTPs. The attribution module of their proposed method implements an ML classifier on the set of TTPs seen during the attack and performs attribution classification to find linked threat groups. They extract TTPs from heterogeneous threat alerts, i.e., IDS/IPS, SIEM, and firewall. Their data modeling stage completes the partial TTP view by predicting the next attack stage and completing the missed TTP based on the cyber kill-chain phase across the MITRE tactics.

Xiao et al.~\cite{xiao2024apt} introduce an APT actor attribution method named APT-MMF, focusing on multimodal and multilevel feature fusion from CTI reports. This method employs multilevel heterogeneous graph attention networks that integrate IoC type-level, meta path-based neighbor node-level, and meta path semantic-level attention to learn deeply hidden features of APT report. Further, they extract and combine three features: attribute types, natural language text, and topological relationships. For natural language text features, they employ BERT encoding~\cite{vaswani2017attention}, and for topological relationship features, they utilize Node2vec~\cite{grover2016node2vec} encoding, and for attribute type features, they use ID encoding and Ordinal encoding. Further, they implement GNN to perform attribution.

\begin{table*}[!hb]
    \centering
    \caption{Comparison of automated attribution methods, their features, and the algorithms employed in each method}
    \label{tab:attributionmethodcomp}
    \begin{tabularx}{\textwidth}{p{0.7cm}p{0.7cm}p{1.8cm}p{1.7cm}XXp{3.5cm}}
    \hline
     \textbf{Paper} & \textbf{Year} & \textbf{Artifact Type} & \textbf{Method Type} & \textbf{Chosen Feature} & \textbf{Feature Transformation} & \textbf{Algorithm}\\
     \hline
     ~\cite{rosenberg2017deepapt} & $2017$ & Behavioral Artifacts & Malware-based & Words of cuckoo sandbox reports & One-Hot encoding  & DNN
     \\
     \\
     ~\cite{perry2019no} & $2019$ & OSINT Artifacts & Threat Report-based & Words of threat intelligence reports & SMOBI word embeddings & XGBoost
     \\
     \\
     ~\cite{noor2019machine} & $2019$ & Behavioral Artifacts & Attack Pattern-based & TTPs & One-Hot encoding  & DNN
     \\
     \\
     ~\cite{haddadpajouh2020mvfcc} & $2020$ & Behavioral Artifacts & Malware-based & OpCode, ByteCode, System call, and Header & One-Hot, Count, TF-IDF, and Eigen vector of CFG & DT
     \\
     \\
     ~\cite{wu2020grouptracer} & $2020$ & Behavioral and Evidentiary Artifacts & Attack Pattern-based & Time, IP, URL, and TTPs mapped from IoT honeypot logs & Statistical characteristics for each feature types & Hierarchical clustering
     \\
     \\
     ~\cite{naveen2020deep} & $2020$ & OSINT Artifacts & Threat Report-based & Words of threat intelligence reports & SIMVER word embeddings & DNN
     \\
     \\
     ~\cite{xu2021apt} & $2021$ & Behavioral Artifacts & Malware-based & API call sequence & n-gram with Adaboost feature selection & DT, KNN, LightGBM, and XGBoost
     \\
     \\
     ~\cite{wang2021explainable} & $2021$ & OSINT and Behavioral Artifacts  & Threat Report and Malware-based & IR code and string features & Paragraph and bag-of-words vectors & RF and DNN
     \\
     \\
     ~\cite{li2021attribution} & $2021$ & Behavioral Artifacts & Malware-based & Words of cuckoo sandbox reports & TF-IDF with Chi-square for dimensionality reduction & SMOTE and RF
     \\
     \\
     ~\cite{kim2021automatically} & $2021$ & Behavioral Artifacts & Attack Pattern-based & TTPs & Tactics-wise One-Hot encoding & k-means clustering
     \\
     \\
     ~\cite{shin2021art} & $2021$ & Behavioral Artifacts & Attack Pattern-based & TTPs  & Tactics-wise One-Hot encoding & Cosine similarity with weighted tactic vectors
     \\
     \\
     ~\cite{kida2022nation} & $2022$ & Behavioral Artifacts & Malware-based & TLSH, SSDEEP, SDHASH, IMPFUZZY, and  LZJD & NATO encoding & RF, SVM, KNN, and NB
     \\
     \\
     ~\cite{wang2022apt} & $2022$ & Behavioral Artifacts & Malware-based & Control Flow Graph (CFG) & Basic blocks characteristics with PCA for dimensionality reduction & HDBSCAN clustering
     \\
     \\
     ~\cite{lee2022malware} & $2022$ & Behavioral Artifacts & Malware-based & Runtime Modules like Kernel32.dll &  Frequency analysis and  One-Hot  encoding & KNN, SVM, DT, NB, Adaptive and Gradient Boosting
     \\
     \\
     ~\cite{mei2022hybrid} & $2022$ & Behavioral and Evidentiary Artifacts & Heterogeneous Data-based  & Log data & TF-IDF & Multi-class SVM with Particle Swarm Optimization
     \\
     \\
     ~\cite{xu2022hghan} & $2022$ & Evidentiary and OSINT Artifacts & Heterogeneous Data-based  & Web attack heterogeneous graphs  & Node embedding & BiLSTM
     \\
     \\
     ~\cite{edie2023extending} & $2023$  & Behavioral Artifacts &  Attack Pattern-based & MITRE TTP and it's association  & TTP List & Wighted Jaccard Index
     \\
     \\
     ~\cite{noor2023machine} & $2023$ & Behavioral Artifacts & Heterogeneous Data-based  & TTPs and Softwares used & One-Hot encoding & Neural Network
     \\
     \\
      ~\cite{sachidananda2023apter} & $2023$ & Behavioral and Evidentiary Artifacts & Heterogeneous Data-based  & TTPs from IDS/IPS and SIEM/EDR Alerts & TTP list & Sequential classifier
      \\
      \\
       ~\cite{xiao2024apt} & $2024$ & OSINT Artifacts & Threat report-based & Node features of heterogeneous attributed graph schema  & ID encoding, Ordinal encoding, BERT encoding, and Node2vec encoding & GNN
       \\
       \\
       \hline
\end{tabularx}
\end{table*}

Xu et al.~\cite{xu2022hghan} present a method named HGHAN, designed for identifying hacker groups through a heterogeneous graph attention network. The HGHAN framework first extracts relevant node sets and attributes from the Web Attack Heterogeneous Information Network (WAHIN), followed by applying LSTM for attribute feature extraction and Metapath2Vec for structural feature extraction. The fused features are subsequently processed through HAN (Heterogeneous Graph Attention Network) for node embedding, with classification being carried out using BiLSTM to leverage its proficiency in capturing bidirectional dependencies within the feature vectors. They conducted experimental validation using web attack data from \url{Zone-H.org} to demonstrate the performance of HGHAN over other node embedding algorithms.

\begin{figure*}[!ht]
    \centering
    \subfloat[Feature types]
    {\includegraphics[width=2.3in]{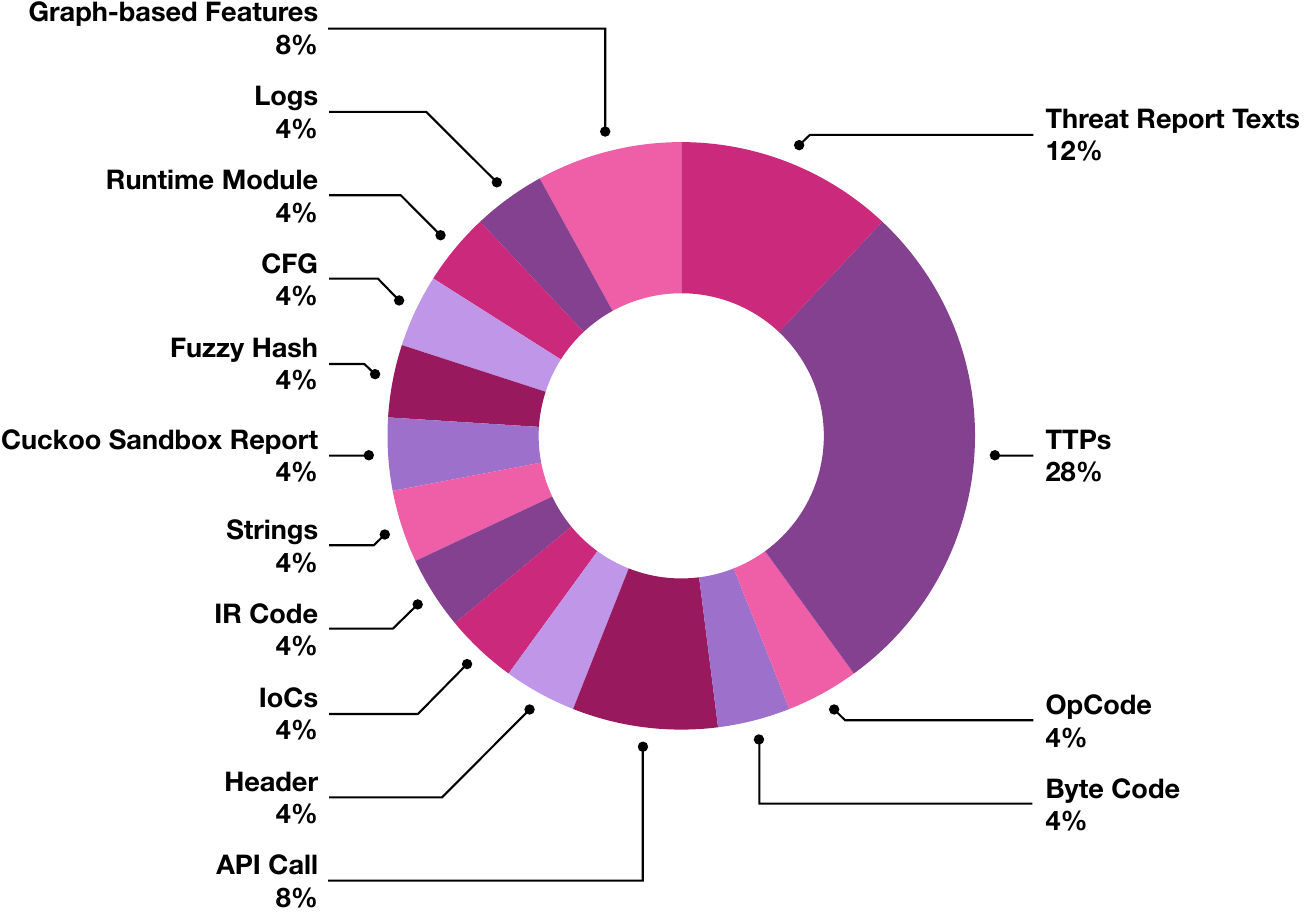}%
    \label{fig:feature_distribution}}%
    \hfil
    \subfloat[Feature transformation methods]
    {\includegraphics[width=2.3in]{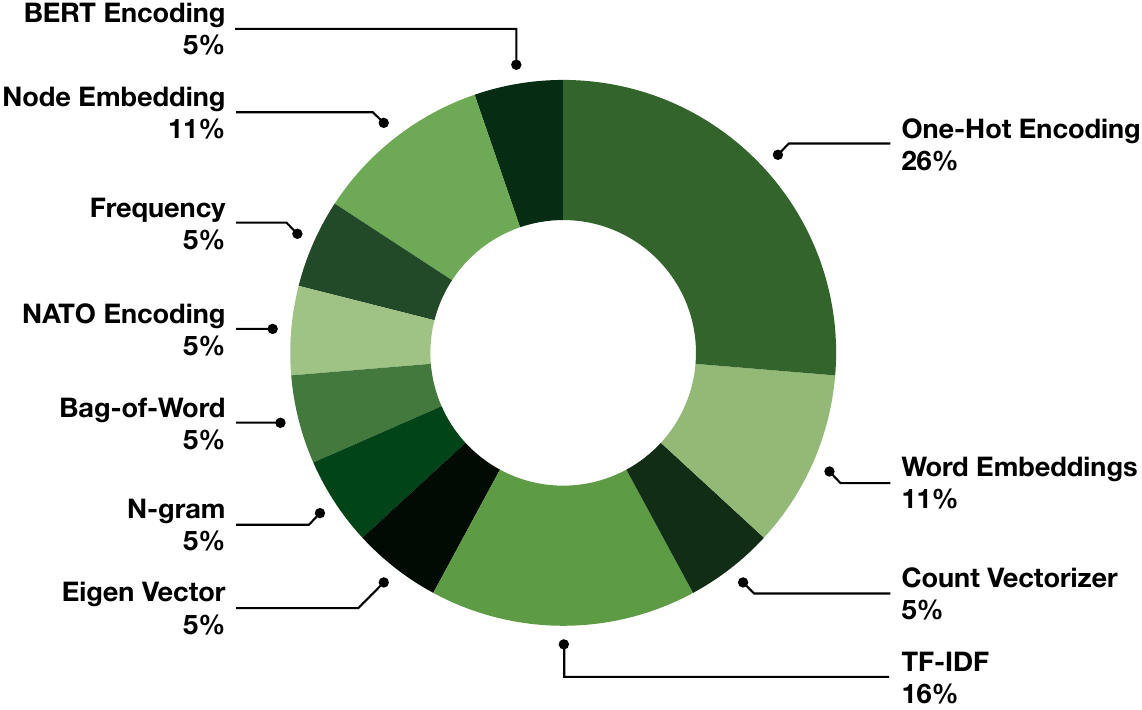}%
    \label{fig:feature_transformation}}%
    \hfil
    \subfloat[Algorithm types]
    {\includegraphics[width=2.4in]{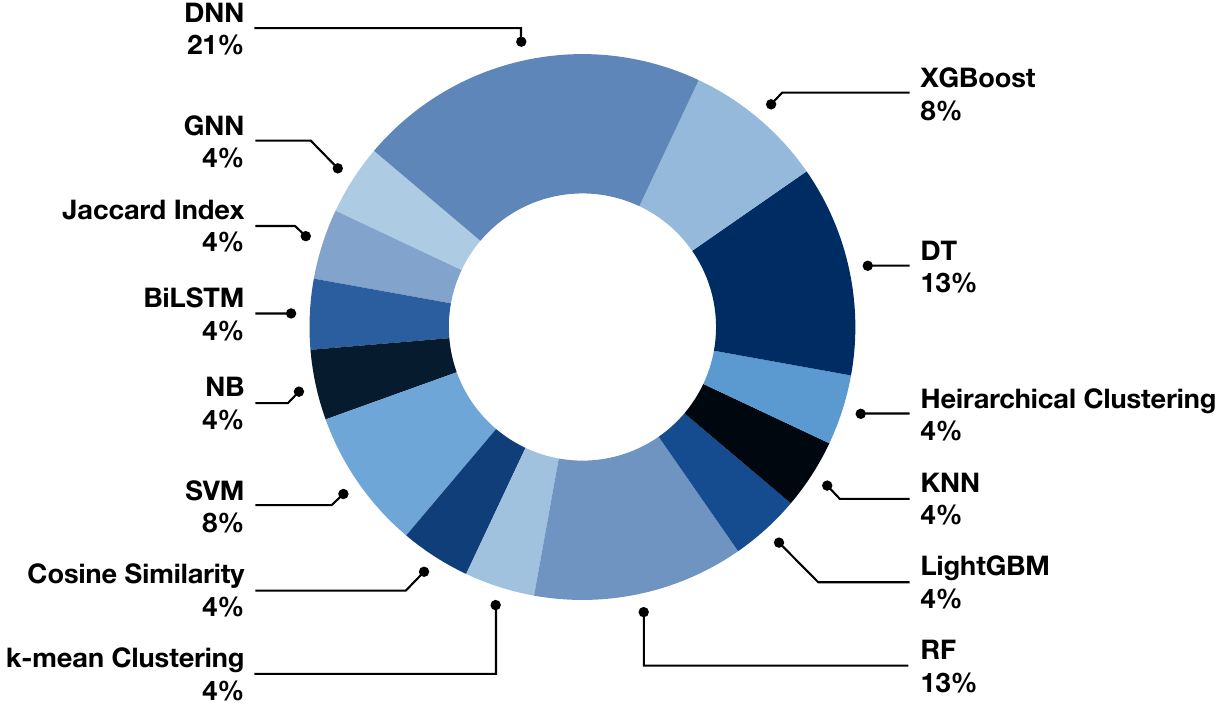}%
    \label{fig:algorithm_distribution}}%
    \caption{Distribution of feature types (a), transformation methods (b), and algorithm types (c).}
    \label{fig:figures}
\end{figure*}

Wu et al.~\cite{wu2020grouptracer} introduce GroupTracer, a framework aimed at identifying and predicting Internet of Things (IoT) attacks by extracting TTP profiles that describe attacker behaviors and identifying potential attacker groups behind complex attacks. GroupTracer captures attack data via IoT honeypots and utilizes the ATT\&CK framework to map and extract TTP profiles automatically. It considers total 18 different features derived from four feature category: TTP profiles, Time, IP, and URL features-to mine potential attack groups. 
They leverage the hierarchical clustering algorithm’s ability to capture the hierarchical structure between clusters. 
\\
\\
A comparison of used features, their transformation, and the algorithm employed by the discussed methods are shown in Table~\ref{tab:attributionmethodcomp}.

For malware-based methods, authors in literature extract the following types of features from malware samples: words from cuckoo sandbox reports, OpCode, ByteCode, API call, Header data, IR (Intermediate Representation) codes, strings, Fuzzy hashes (TLSH, SSDEEP, SDHASH, IMPFUZZY, and LZJD), CFG, and runtime modules. Threat report-based attribution methods extract words from threat intelligence reports as features. Attack pattern-based methods leverage TTPs as a feature for attribution. Meanwhile, heterogeneous data-based attribution methods leverage diverse sets of features like TTPs, software used, and heterogeneous attributed graph schemas. The distribution of feature types used in the literature is shown in Fig.~\ref{fig:feature_distribution}.
Out of all features mentioned in the literature TTPs and threat report texts has been used widely to present attribution method.

For feature transformation, the literature leverages the following different methods: One-Hot encoding, SMOBI word embedding, TF-IDF, Count, Statistical Characteristics, SIMVER word embedding, n-gram method, NATO encoding, Basic block characteristics, frequency analysis, Node embedding, BERT embedding, and Node2vec embedding. The distribution of type of transformation methods has been used in the literature is shown in Fig.~\ref{fig:feature_transformation}. Out of all used transformation methods One-Hot encoding and TF-IDF has been used widely in the literature.

Literature methods employ the following algorithms for classification or clustering: DNN, XGBoost, DT, Hierarchical clustering, KNN, LightGBM, RF, k-means clustering, Cosine similarity, SVM, NB, HDBSCAN (Hierarchical Density-Based Spatial Clustering of Applications with Noise) clustering, BiLSTM, Weighted Jaccard Index, DNN, Sequential classifier, and GNN. The distribution of type of algorithms used in the literature is shown in Fig.~\ref{fig:algorithm_distribution}. Out of all employed classification model DNN, DT, and RF has been used widely in the literature to train attribution model.


\subsection{Discussion}
\label{subsec:discussion}
The discussed papers demonstrate a range of strengths and weaknesses in their approaches to threat attribution. In this section, we provide a detailed analysis of these aspects, thoroughly examining the methodologies and outcomes of each paper.

We review the literature methods comprehensively and find five major data related problems among the available methods, which we discussed below. 

\textbf{(P1)} As malware can be tweaked, distributed, or sold, i.e., given the rise of MaaS (Malware-as-a-Service)~\cite{patsakis2024malware}, it may lead to notable challenges in uncovering the actual origin of a cyber attack. Additionally, malwares' behaviour-based threat attribution may provide an inconsistent view of threat actors when the same threat actor employs numerous malware in various campaigns. \textbf{(P2)} Malware-based threat attribution can be misleading in the case of collaborative and multi-layered attacks, as these attacks do not rely on a single instance of malware. \textbf{(P3)} For the threat-report-based attribution methods, authors employ threat report words as features for classification in the literature. Analysing the words of two threat reports shows the similarity between threat reports rather than the similarity between the threat actor’s behavior or modus operandi. Comparing words for document similarity has been studied widely in the natural language processing domain~\cite{sedding2004wordnet,selva2021review}. \textbf{(P4)} Given P3, such methods need to reverse engineer and identify words contributing to the attribution or profiling of a specific threat actor. \textbf{(P5)} Attack pattern-based methods leverage TTPs in the literature. TTP of a threat group can evolve over time~\cite{RussianTTPEvolve}, and TTP-based threat attribution can miss-attribute in such cases. Therefore, TTP-based attribution methods demand continuous updates according to the threat group’s latest modus operandi. Also, threat actors may employ false flags by employing other groups of TTPs to disrupt the attribution method or hide their traces~\cite{bartholomew2016wave,skopik2020under}. 

\begin{table*}[!hb]
    \centering
    \caption{Comparison of attribution methods and their performance}
    \label{tab:methodperformancecomp}
    \begin{threeparttable}
    \begin{tabularx}{\textwidth}{p{0.5cm}p{1cm}p{1cm}p{1cm}p{1.1cm}p{0.7cm}p{0.6cm}p{0.6cm}p{0.9cm}X}
    \hline
    \textbf{Paper} & \textbf{Accuracy (\%)} & \textbf{Precision (\%)} & \textbf{Recall (\%)} & \textbf{F1-score (\%)} & \textbf{Context-aware} & \textbf{Source Code} & \textbf{Open World} & \textbf{Granular} & \textbf{Comments}
    \\
    \hline
    ~\cite{rosenberg2017deepapt} & $94.60$ & $-$ & $-$ & $-$ & \quad \ding{55} & \quad \low & \quad \ding{55} & \quad \ding{55} & \begin{itemize}[nosep, leftmargin=*] 
         \vspace{-0.2cm}
            \item Attributes at the country level rather than the threat group level
            \item Word semantics of cuckoo report reflect malware functionality rather than threat actor behavior
            \item As the method is leveraging malware sample to attribute threat groups, it faces problems \textbf{P1} and \textbf{P2} as discussed in the Section~\ref{subsec:discussion}
        \end{itemize}
    \\
    ~\cite{perry2019no} & $58.40$ & $55.00$ & $52.40$ & $-$ &  \quad \ding{55} & \quad \medium & \quad \ding{51} & \quad \ding{51} & \begin{itemize}[nosep, leftmargin=*] 
        \vspace{-0.2cm}
            \item Their feature vector method, SMOBI,  focuses on word frequency and similarity, ignoring semantics and context
            \item As the method is leveraging words present in the threat report to attribute threat groups, it faces problems \textbf{P3} and \textbf{P4} as discussed in the Section~\ref{subsec:discussion}
        \end{itemize}
    \\
    ~\cite{noor2019machine} & $94.00$ & $90.00$ & $89.00$ & $89.00$  &  \quad \ding{55} & \quad \low & \quad \ding{55} & \quad \ding{51}  & \begin{itemize}[nosep, leftmargin=*] 
         \vspace{-0.2cm}
            \item  Model is trained on the highly imbalanced and limited dataset
            \item Their TTP extraction method is based on semantic  similarity rather than the contextual meaning of sentences
            \item The feature transformation method considers only the existence of TTPs and ignores contextual information such as when and how TTP has been used
        \end{itemize}
    \\
    ~\cite{haddadpajouh2020mvfcc} &  $95.00$ & $-$ & $-$ & $-$ & \quad \ding{55} & \quad \medium & \quad \ding{55} & \quad \ding{51}  & \begin{itemize}[nosep, leftmargin=*] 
         \vspace{-0.2cm}
            \item Dataset used is limited to five APT groups; lacks diversity. A larger and more diverse dataset would strengthen the findings
            \item Claims significant improvements but lacks comparisons with the latest state-of-the-art methods in malware-based attribution
            \item As the method is leveraging malware sample to attribute threat groups, it faces problems \textbf{P1} and \textbf{P2} as discussed in the Section~\ref{subsec:discussion}
        \end{itemize}
    \\
    ~\cite{wu2020grouptracer} & $-^\star$ & $-^\star$ & $-^\star$ & $-^\star$ & \quad \ding{55} & \quad \low & \quad \ding{55} & \quad \ding{51} & \begin{itemize}[nosep, leftmargin=*] 
         \vspace{-0.2cm}
         \item Dataset used is limited to specific honeypots (UPnP-SOAP and Netis router backdoor)
         \item Proposed method compare with two traditional clustering method but lack the comparison with state-of-the-art attribution methods
    \end{itemize}
    \\
    ~\cite{naveen2020deep} & $86.50$ & $95.40$ & $83.30$ & $87.90$ & \quad \ding{55} & \quad \low & \quad \ding{55} & \quad \ding{51} & \begin{itemize}[nosep, leftmargin=*] 
         \vspace{-0.2cm}
             \item The feature transformation method, SIMVER, focuses on word semantics but ignores the contexts of sentences
             \item As the method is leveraging words present in the threat report to attribute threat groups, it faces problems \textbf{P3} and \textbf{P4} as discussed in the Section~\ref{subsec:discussion}
        \end{itemize}
    \\
    ~\cite{xu2021apt} & $93.53$ & $89.65$ & $84.72$ & $86.62$ & \quad \ding{51} & \quad \low & \quad \ding{55} & \quad \ding{51} & \begin{itemize}[nosep, leftmargin=*] 
         \vspace{-0.2cm}
         \item Computationally intensive with large feature sets. Despite selecting the top 400 features, scalability remains a concern with larger dataset
         \item Detail comparison with various ML classifiers but lacks comparison with state-of-the-art malware-based attribution methods
         \item As the method is leveraging malware sample to attribute threat groups, it faces problems \textbf{P1} and \textbf{P2} as discussed in the Section~\ref{subsec:discussion}
         \end{itemize}
    \\
    ~\cite{wang2021explainable} & $97.87$ & $-$ & $-$ & $-$ & \quad \ding{55} & \quad \low & \quad \ding{55} & \quad \ding{51} & \begin{itemize}[nosep, leftmargin=*] 
         \vspace{-0.2cm}
             \item Rely on static features for the decision which is challenging in the case of obfuscated samples. The obfuscation is common in APT types of attacks
             \item As the method is leveraging malware sample to attribute threat groups, it faces problems \textbf{P1} and \textbf{P2} as discussed in the Section~\ref{subsec:discussion}
            \end{itemize}
    \\
    \hline
    \multicolumn{10}{>{\footnotesize\itshape}r}{Continue on the next page}
\end{tabularx}
\begin{tablenotes}
        \item[] \high\  Source code is available, \medium\  Only data is available, \low\  No source code is available
        \item[] $-$ Not Available
        \item [] $-^\star$ Calinski–Harabasz score: $3416.9311$, Silhouette Coefficient: $0.5389$, Davies–Bouldin: $0.7367$
        \end{tablenotes}
        \end{threeparttable}
\end{table*}

\begin{table*}[!h]
    \ContinuedFloat
    \centering
    \caption{Comparison of attribution methods and their performance}
    \label{tab:methodperformancecomp}
    \begin{threeparttable}
    \begin{tabularx}{\textwidth}{p{0.5cm}p{1cm}p{1cm}p{1cm}p{1.1cm}p{0.7cm}p{0.6cm}p{0.6cm}p{0.9cm}X}
    \hline
    \textbf{Paper} & \textbf{Accuracy (\%)} & \textbf{Precision (\%)} & \textbf{Recall (\%)} & \textbf{F1-score (\%)} & \textbf{Context-aware} & \textbf{Source Code} & \textbf{Open World} & \textbf{Granular} & \textbf{Comments}
    \\
    \hline
    ~\cite{li2021attribution} & $-$ & $85.00$ & $79.10$ & $79.40$ & \quad \ding{55} & \quad \low & \quad \ding{55} & \quad \ding{51} & \begin{itemize}[nosep, leftmargin=*] 
         \vspace{-0.2cm}
         \item No evaluation to demonstrate the data generated by SMOTE aligns with base data and not contributing towards biases
         \item No comparison with state-of-the-art malware-based threat attribution methods
         \item As the method is leveraging malware sample to attribute threat groups, it faces problems \textbf{P1} and \textbf{P2} as discussed in the Section~\ref{subsec:discussion}
         \end{itemize}
    \\
    ~\cite{kim2021automatically} & $-$ & $91.48$ & $95.14$ & $93.27$ & \quad \ding{55} & \quad \low & \quad \ding{55} & \quad \ding{51} & \begin{itemize}[nosep, leftmargin=*] 
         \vspace{-0.2cm}
         \item IoC pairing method can mislead in false flags as chosen IoCs like C\&C Platform, their IP and malware types, can be modified or changed in APT attacks
         \item No comparison with state-of-the-art malware-based threat attribution methods
         \item As the method is leveraging malware sample to attribute threat groups, it faces problems \textbf{P1} and \textbf{P2} as discussed in the Section~\ref{subsec:discussion}
         \end{itemize}
    \\
    ~\cite{kida2022nation} & $89.00$ & $-$ & $-$  & $88.00$ & \quad \ding{55} & \quad \low & \quad \ding{55} & \quad \ding{51} & \begin{itemize} [nosep, leftmargin=*]
         \vspace{-0.2cm}
            \item The authors reported their method struggles to surpass the state-of-the-art, aiming to understand how fuzzy hashing aids in identifying nation-state actors
            \item As the method is leveraging malware sample to attribute threat groups, it faces problems \textbf{P1} and \textbf{P2} as discussed in the Section~\ref{subsec:discussion}
         \end{itemize}
    \\
    ~\cite{wang2022apt} & $-^\circledast$ & $-^\circledast$ & $-^\circledast$ & $-^\circledast$ & \quad \ding{55} & \quad \low & \quad \ding{55} & \quad \ding{51} & \begin{itemize}[nosep, leftmargin=*] 
         \vspace{-0.2cm}
         \item Relies on static features, which can be less effective against advanced sophisticated malware which employs various obfuscation methods
         \item As the method is leveraging malware sample to attribute threat groups, it faces problems \textbf{P1} and \textbf{P2} as discussed in the Section~\ref{subsec:discussion}
         \end{itemize}
    \\
    ~\cite{mei2022hybrid} & $93.96$  & $93.97$  & $-$ & $93.94$ & \quad \ding{55} & \quad \low & \quad \ding{55} & \quad \ding{51} & \begin{itemize}[nosep, leftmargin=*] 
         \vspace{-0.2cm}
         \item No comparison with state-of-the-art malware-based threat attribution methods
         \item As the method is leveraging malware sample to attribute threat groups, it faces problems \textbf{P1} and \textbf{P2} as discussed in the Section~\ref{subsec:discussion}
         \end{itemize}
    \\
    ~\cite{edie2023extending} & $95.70$& $-$ & $-$ & $-$ &  \quad \ding{55} & \quad \high & \quad \ding{55} & \quad \ding{51} & \begin{itemize}[nosep, leftmargin=*] 
         \vspace{-0.2cm}
         \item Reporting only accuracy is insufficient; additional metrics like precision, recall, and F1-score are needed for a comprehensive evaluation.
         \item Lacks a thorough comparison with existing state-of-the-art methods for APT attribution
         \item As the method is leveraging TTPs as feature for attribution, it faces problem \textbf{P5} as discussed in the Section~\ref{subsec:discussion}
         \end{itemize} \\
     ~\cite{noor2023machine} & $94.88$ & $93.95$ & $94.88$ & $94.00$ & \quad \ding{55} & \quad \medium & \quad \ding{55} & \quad \ding{51} & \begin{itemize}[nosep, leftmargin=*] 
         \vspace{-0.2cm}
         \item Lacks a thorough comparison with existing state-of-the-art methods for cyber threat attribution
         \item While preparing dataset, the process of synthesis of the high-level IoC dataset is not explained
         \end{itemize}
    \\
    ~\cite{sachidananda2023apter} & $97.30$ & $96.00$ & $97.00$ & $97.00$ & \quad \ding{51} & \quad \low & \quad \ding{55} & \quad \ding{51} & \begin{itemize}[nosep, leftmargin=*] 
         \vspace{-0.2cm}
         \item While preparing dataset, mapping SIEM/EDR alerts to MITRE ATT\&CK TTP is done by a security tool but there is no mention of the tool name
         \item The performance of the method is depends on the mapping efficiency of chosen security tool using which data is prepared
         \item As the method is leveraging TTPs as feature for attribution, it faces problem \textbf{P5} as discussed in the Section~\ref{subsec:discussion}
         \end{itemize}
    \\
    ~\cite{xiao2024apt} & $-$ & $-$ & $-$ & $70.51$ & \quad \ding{51} & \quad \low  & \quad \ding{55} & \quad \ding{51} & \begin{itemize}[nosep, leftmargin=*] 
         \vspace{-0.2cm}
         \item Lacks a thorough comparison with existing state-of-the-art methods for cyber threat attribution
         \item The accuracy and efficiency of analytical calculations depend on the lengths and complexity levels of metapaths. There is no evaluation presented regarding computational complexity
         \end{itemize}   \\
    \hline
    \multicolumn{10}{>{\footnotesize\itshape}r}{Continue on the next page}
    \end{tabularx}
    \begin{tablenotes}
        \item[] \high\  Source code is available, \medium\  Only data is available, \low\  No source code is available
        \item[] $-$ Not Available
        \item[] $-^\circledast$ AUC (Area Under the Curve): $93.6\%$
        \end{tablenotes}
        \end{threeparttable}
\end{table*}

\begin{table*}[!h]
    \ContinuedFloat
    \centering
    \caption{Comparison of attribution methods and their performance}
    \label{tab:methodperformancecomp}
    \begin{threeparttable}
    \begin{tabularx}{\textwidth}{p{0.5cm}p{1cm}p{1cm}p{1cm}p{1.1cm}p{0.7cm}p{0.6cm}p{0.6cm}p{0.9cm}X}
    \hline
    \textbf{Paper} & \textbf{Accuracy (\%)} & \textbf{Precision (\%)} & \textbf{Recall (\%)} & \textbf{F1-score (\%)} & \textbf{Context-aware} & \textbf{Source Code} & \textbf{Open World} & \textbf{Granular} & \textbf{Comments}
    \\
    \hline
    ~\cite{xu2022hghan} & $98.00$ & $97.00$ & $98.00$ &  $97.00$ & \quad \ding{51} &  \quad \low & \quad \ding{55} & \quad \ding{55} & \begin{itemize}[nosep, leftmargin=*] 
         \vspace{-0.2cm}
         \item  No comparison with state-of-the-art methods
         \item Dataset used is limited to web-based application, i.e., Web-Hacking dataset
         \end{itemize}
    \\
    ~\cite{lee2022malware} & $87.10$ & $89.00$ & $87.00$ & $87.0$ & \quad \ding{55} & \quad \low & \quad \ding{55} & \quad \ding{51} & \begin{itemize}[nosep, leftmargin=*] 
         \vspace{-0.2cm}
         \item Dataset used is limited to only three APT group
         \item As the method is leveraging malware sample to attribute threat groups, it faces problems \textbf{P1} and \textbf{P2} as discussed in the Section~\ref{subsec:discussion}
         \end{itemize}
    \\
    \hline
    \end{tabularx}
    \begin{tablenotes}
        \item[] \high\  Source code is available, \medium\  Only data is available, \low\  No source code is available
        \item[] $-$ Not Available
        \end{tablenotes}
        \end{threeparttable}
\end{table*} 

Further, we provide critical comments on the current literature methods and their reported performance in the respective studies in Table~\ref{tab:methodperformancecomp}.
The discussed critical comments appeal to evaluate literature methods and future guidance for new research dimensions in threat attribution. While reviewing these literatures, we also find that the majority of the literature suffers from the following limitations:
\begin{enumerate}[label=(\alph*)]
    \item \textbf{Comparison with state-of-the-art methods:} Majority of the literature lacks the comparison with the state-of-the-art attribution method. The reported performance of literature methods are good, but a comparison with state-of-the-art methods in the same environment and common dataset can establish a comprehensive comparison and prove the efficiency of the proposed methods.
    \item \textbf{Limited threat group implementation:} The number of threat groups researchers use to implement their methods is limited, with very few APT groups. For example, the method of~\cite{haddadpajouh2020mvfcc} is tested on only five APT group samples, and the method of~\cite{shin2021art} is tested on only three APT group samples. Such limited environment testing raises concerns regarding method effectiveness for diverse and complex real-world data.
    \item \textbf{Lacks open-world solutions:} APT groups evolve over time, and several threat groups emerge nowadays. It raises concerns regarding models trained on close-world solutions. Close-world solutions attribute threat groups to only what they have seen during training. The emergence of threat groups leads to misattribution when the given sample belongs to newly emerged groups whose sample has not been seen by the model during training. As a result, the trained model is attributed to a wrong threat group. 
    Our review also finds that only~\cite{perry2019no} introduced the open-world solution, i.e., capability of their method for identifying unknown or newly emerged threat actors.
    \item \textbf{Lacks contextual information:}  The current attribution literature is majorly based on the existence of a feature rather than leveraging its contextual information. It is crucial to see how an artifact is being used during the attack to connect the dots with historical threat groups’ behavior~\cite{irshad2024context}. For example, TTP X is used to gain the initial access, and subsequently TTP Y is used for privilege escalation. This sequential information may hint at more aligned behavior rather than just matching the appearance of the TTPs individually. Most of the literature leverages the One-Hot encoding and TF-IDF feature transformation method to convert selected features into vectors, which does not  consider contextual information. Some of the work has leveraged contextual information of the artifacts to perform attribution~\cite{xu2021apt,xu2022hghan,sachidananda2023apter,xiao2024apt}. Using context-aware feature transformation methods or consideration of contextual information can assist researchers in developing attribution methods that make informed decisions.
    \item \textbf{Granularity level:} Granularity in attribution can significantly influence the effectiveness and reliability of the attribution process. It refers to the level of detail and precision with which cyber attacks are attributed to their perpetrators. According to 4C model~\cite{steffens2020attribution}, the granularity of attribution is divided into four levels: Technical, Motivation, Country, and Organization/Person. Technical attribution involves identifying and analyzing technical artifacts like malware signatures and IP addresses. Motivation attribution focuses on understanding the underlying reasons for the attack, such as financial gain, espionage or disruption. Country attribution aims to determine the geographical origin of the attackers by examining factors like language settings and geopolitical interests. The most detailed level, Organization/Person attribution, involves identifying specific organizations or individuals responsible for the attack, often requiring extensive investigation and evidence gathering. Each level provides a deeper understanding of the attack, contributing to a comprehensive attribution process. Literature like~\cite{rosenberg2017deepapt} focuses on attributing only nation rather than the specific threat group. The attribution demands uncovering as many specific details as possible about the threat actor. Therefore, fine-grained attribution allows for more tailored and effective defensive measures, enhanced intelligence sharing, and informed strategic decision-making.
    \item \textbf{Performance metrics:} The performance measured in the current literature is dispersed. Some of the current methods~\cite{rosenberg2017deepapt,haddadpajouh2020mvfcc,wang2021explainable,edie2023extending} compare the efficiency of literature models based on accuracy measures, Whereas some of the literature focus on other metrics like precision, recall, or f1-score, such as~\cite{wu2020grouptracer,wang2022apt}. The goal for the specific case of attribution is to get better link between given artifact and associated threat group. Along with this, incorrect attribution can result in misdirected cybersecurity efforts, wasted resources, and possible diplomatic conflicts, weaken the credibility of the attribution process, escalate tensions, and may lead to incorrect strategic decisions~\cite{BlueGoatCyber_Attribution}. Therefore, the false positive needs to be minimized, i.e., precision is more significant than other performance metrics in threat attribution. 
\end{enumerate}

\section{Challenges}
\label{sec:challenges}
Attribution itself is a challenging and complex task and requires diverse consideration during the process. Additionally, APT attacks’ sophisticated, multifaceted, multi-staged, and dynamic nature makes it more difficult. 
The challenges observed by APT attribution methods are discussed in this section.

\subsection{Lack of comprehensive and structured data}
\label{subsec:lackofdata}
A significant challenge in APT attack attribution is the lack of comprehensive and structured datasets. The available threat intelligence is often incomplete or fragmented, making it difficult to form a complete picture of the threat landscape. This fragmented data can lead to partial or incorrect conclusions, as investigators might miss critical information necessary for accurate attribution. Our presented artifact taxonomy serves as a foundational step toward creating structured data for attribution. Researchers can utilize this taxonomy to identify and collect the relevant data types necessary for attribution. Furthermore, the quality and reliability of threat intelligence data vary widely, with some sources providing high-fidelity data while others contribute less reliable information. This variability further complicates the accuracy of attribution efforts, as inconsistent and unreliable data can result in incorrect attributions~\cite{li2024advanced}. Without robust and structured datasets, the attribution process remains severely hindered, potentially leading to flawed assessments and ineffective responses to cyber threats.
\vspace{-0.3cm}
\subsection{False flags}
\label{subsec:falseflag}
APT groups may deliberately leave false clues, known as false flags, to mislead investigators about their true identity~\cite{skopik2020under,egloff2020public,alrabaee2017feasibility}. This deceptive tactic involves using tools, techniques, and methods typically associated with other groups or countries~\cite{bartholomew2016wave,pitropakis2018enhanced}. By adopting these recognizable signatures, they aim to create confusion and misdirection, making it appear that another entity is responsible for the attack. This strategy complicates the attribution process and increases the likelihood that investigators will draw incorrect conclusions about the origin and perpetrators of the cyber assault. False flags are the sophisticated method to obscure the true actors behind an attack, highlighting the complexity and challenges in accurate cyber threat attribution.
\vspace{-0.3cm}
\subsection{Lack of robustness against coordinated attack}
\label{subsec:robustness}
Attribution efforts face significant challenges due to the attack-latent relationships between APT groups. Some APT groups collaborate against common victims to conduct joint operations and strive toward mutual strategic goals~\cite{li2024advanced}. This collaboration complicates the attribution process, blurring the lines between different groups’ activities. When APT groups work together, it becomes challenging to determine which group is responsible for a specific attack. The overlapping use of shared resources and coordinated strategies can mislead investigators, making it difficult to pinpoint the attack’s origin and attribute it to the correct entity. This complexity underscores the need for advanced and nuanced attribution methods capable of disentangling these collaborative efforts.

\subsection{Shared tools}
\label{subsec:sharedtools}
Publicly available tools are commonly used by many APT groups~\cite{gray2024identifying}, making it challenging to differentiate between actors solely based on tool usage. For example, both APT29 and Kimsuky utilize PsExec, though they are reportedly sponsored by different nation-states~\cite{gray2024identifying}. These shared tools and techniques among various threat actors further blur the lines, making it difficult to attribute an attack to a specific group solely based on technical indicators. When multiple groups use the same tools, it becomes challenging to identify unique signatures or patterns that could definitively link an attack to a particular group. This widespread use of shared resources complicates the attribution process, as investigators cannot rely on tool usage as a clear differentiator. Thus, the shared nature of these tools necessitates more sophisticated methods to attribute cyber attacks to the correct perpetrators accurately.

\vspace{-0.25cm}
\subsection{Anonymization}
\label{subsec:anonymization}
APT groups often employ sophisticated anonymization techniques, such as Tor, Virtual Private Networks (VPNs), and proxy servers, to mask their actual IP addresses and geographical locations~\cite{boebert2010survey}. Using these methods, they can effectively conceal their origins, making it exceedingly difficult for investigators to pinpoint the trustworthy source of an attack. These deliberate misdirections complicate the attribution process, as they create layers of obfuscation that need to be penetrated to identify the attackers accurately. The presence of such anonymization necessitates advanced forensic techniques and extensive analysis to discern the real origin of the threat. Automated systems often struggle to achieve this level of performance, and highlights the importance of advanced analytical methods in attributing sophisticated cyber attacks.

\subsection{Interdisciplinary nature}
\label{subsec:interdisciplinarynature}
The interdisciplinary nature of cyber threat attribution adds another layer of complexity. Accurate attribution requires integrating technical expertise in cybersecurity, malware analysis, and network forensics with non-technical insights from geopolitics, sociology, and intelligence analysis~\cite{swate2024analysis}. This interdisciplinary approach is necessary to understand an attack’s broader context and motivations. However, achieving such integration is challenging and requires collaboration between stakeholders, including private companies, governments, and international organizations. 

\subsection{Legal and ethical implications}
\label{subsec:legalandethical}
The legal and ethical considerations of cyber threat attribution are essential and must be diligently addressed~\cite{tsagourias2020cyber}. Collecting and analyzing data for attribution can infringe on privacy rights and raise significant ethical concerns~\cite{jaafar2020demystifying}. Incorrect or premature attribution can lead to severe legal and diplomatic repercussions, potentially escalating conflicts and causing harm to innocent parties. These considerations necessitate a cautious and methodical approach along with automated systems. Balancing the need for accurate and timely attribution with the ethical and legal ramifications adds to the overall difficulty of the task.

\section{Open Research Problems}
\label{sec:openresearchproblems}

\begin{figure}
    \centering
    \includegraphics[width=\linewidth]{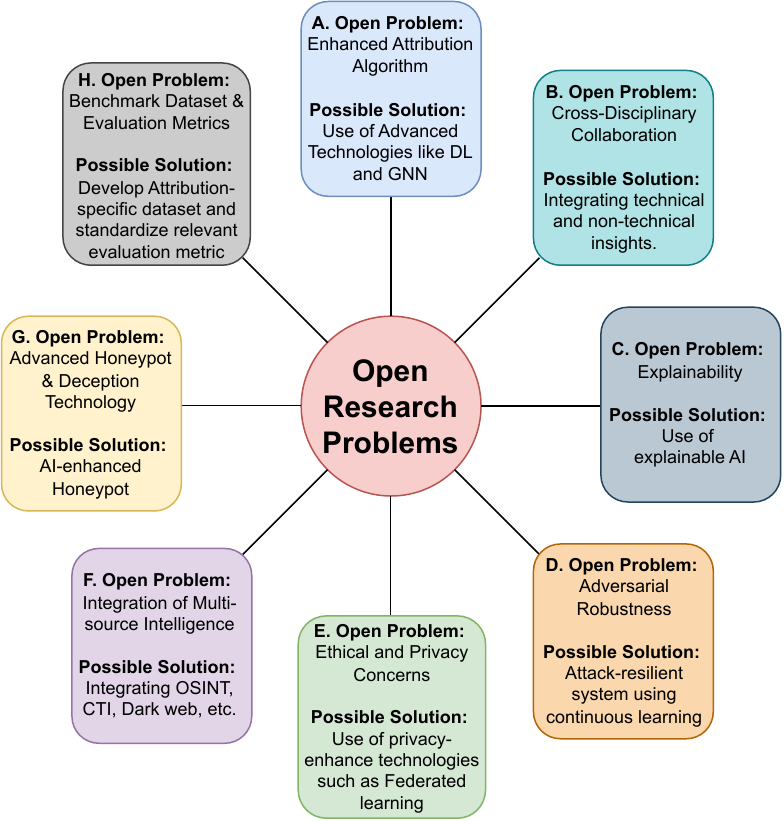}
    \caption{Open research problems and possible future research directions}
    \label{fig:openresearchproblem}
\end{figure}

The difficulty of automated cyber threat attribution presents numerous open research opportunities and future research directions. As cyber threats evolve in sophistication and complexity, there is a growing need for innovative approaches and technologies to improve the accuracy and reliability of threat attribution. Based on our survey, we present some critical open research problems where future research can significantly advance APT attribution as shown in Fig.~\ref{fig:openresearchproblem} and discussed in this section.

\subsection{How does one enhance the current automated APT attribution methodologies?}
\label{subsec:enhancedattributionalgo}
Automated cyber threat attribution presents numerous research opportunities that promise to advance the field of cybersecurity significantly. Enhancing attribution algorithms through sophisticated ML and AI techniques remains a top priority. By incorporating advanced models such as GNNs, researchers can better recognize patterns within complex cyber attack data, leading to more accurate and efficient attribution. These models excel at identifying intricate relationships and connections within the data that traditional methods might miss. Additionally, context-aware attribution methods are being developed to improve accuracy in considering the broader context of an attack, such as geopolitical factors and historical attack patterns. By integrating these advanced techniques, automated attribution systems can become more robust and reliable, offering a powerful tool for cybersecurity professionals to identify and respond to threats.

\subsection{How can insights from diverse fields be effectively integrated into automated attribution systems to enhance their robustness and depth?}
\label{subsec:crossdisciplinaricolab}
Cyber threat attribution significantly benefits from cross-disciplinary collaboration, combining insights from various fields such as cybersecurity, geopolitics, and open-source and signal intelligence analysis. This integrated approach ensures that technical data is enriched with a contextual understanding, providing a more comprehensive view of the threats. Researchers can explore automated frameworks and methodologies to foster such collaboration, creating systems where experts from different disciplines can contribute their unique perspectives. This interdisciplinary approach can lead to more robust and nuanced attribution conclusions, allowing for a deeper understanding of cyber attacks’ motives, methods, and implications. By leveraging the strengths of diverse fields, the attribution process can become more accurate and insightful, ultimately enhancing overall cybersecurity efforts.

\subsection{How can we create explainable and transparent attribution algorithms that provide clear and understandable reasoning for their decisions to gain trust and facilitate adoption?}
\label{subsec:explainablity}
The transparency and trustworthiness of AI models used in threat attribution are essential focus areas. Developing explainable AI models is crucial, as these models need to provide clear and understandable reasoning for their decisions. This transparency is vital for gaining user trust, as stakeholders must understand how and why certain conclusions are reached. Explainable AI can help users feel more confident in the technology and ensure its adoption in critical applications. By making AI models more interpretable, developers can bridge the gap between complex algorithms and human understanding to facilitate better decision-making and foster greater acceptance and reliance on AI-driven threat attribution systems.

\subsection{How can we design attribution algorithms robust to adversarial attacks used by sophisticated threat actors to mislead investigators?}
\label{subsec:adversarialrobust}
As cyber threats evolve, attribution models must enhance their robustness against adversarial attacks. It involves designing models that are resilient to attempts at misleading attribution efforts. Attackers often use sophisticated methods to deceive these models, so creating attack-resilient systems is essential. Additionally, implementing continuous learning mechanisms allows these models to adapt to the ever-changing threat landscape. This adaptive capability ensures that automated attribution systems remain effective over time, maintaining their ability to accurately identify and attribute cyber threats despite the evolving tactics of adversaries. Robust and adaptable models are crucial to sustain the reliability and performance of cyber threat attribution in a dynamic environment.

\subsection{How to develop privacy-preserving techniques to protect sensitive data in attribution processes?}
\label{subsec:ethicalandpriv}
Ethical and privacy considerations are critical in the development of future attribution technologies. Researchers must create privacy-preserving techniques to ensure that attribution processes do not infringe on user privacy. It involves implementing methods that protect sensitive data while allowing accurate attribution. Additionally, establishing ethical guidelines is essential to govern the responsible use of these technologies. These guidelines will help ensure that attribution efforts are conducted in a manner that respects individual rights and adheres to legal and moral standards. By addressing ethical and privacy concerns, developers can foster trust and acceptance of attribution technologies, ensuring their responsible and effective application in cybersecurity.

\subsection{How can diverse data sources be effectively integrated and harmonized to improve the APT attribution?}
\label{subsec:integratemultisourceintel}
Integrating multiple intelligence sources, such as OSINT and traditional CTI, can significantly enhance the understanding of threat actors and their activities. OSINT, which includes publicly available information from various online platforms, can complement traditional intelligence methods by providing additional context and insights. Additionally, mining dark web data for signs of attack planning and correlating it with real-world incidents offers a rich, untapped potential. This approach can reveal deeper insights into cybercriminals’ motives and methods, allowing for a more comprehensive and nuanced view of the threat landscape. By leveraging OSINT and dark web data, security professionals can develop a more detailed and accurate picture of cyber threats, improving their ability to anticipate and respond to attacks.

\subsection{What methodologies can be developed to improve the ability of honeypots to detect and capture malicious activities and accurately identify the perpetrators behind these attacks?}
\label{subsec:advancedhoneypotdeceptiontech}
Advanced honeypots and deception techniques are emerging as promising research areas in cybersecurity. AI-enhanced honeypots, which can analyze captured attack data and automatically attribute it to specific threat actors, offer significant benefits by providing valuable intelligence with minimal human intervention. These sophisticated honeypots detect and capture malicious activity and help identify the perpetrators behind the attacks. Additionally, developing new deception strategies can further improve the accuracy of attribution data. Security researchers can gather detailed insights into threat actors’ behavior, techniques, and tactics by misleading attackers and studying their responses. These advanced methods enhance our understanding of cyber threats and improve the effectiveness of attribution efforts.

\subsection{How do we benchmark datasets and evaluation metrics to rigorously assess and compare the performance of different automated attribution algorithms?}
\label{subsec:benchmarkdatasetandevalmetric}
Benchmark datasets and evaluation metrics are essential for advancing the field of cyber threat attribution. Creating standardized benchmark datasets enable researchers to assess and compare different attribution methods rigorously. These datasets provide a consistent basis for testing, ensuring that new techniques are evaluated against a common standard. Evaluation metrics, particularly those focused on precision, are crucial for measuring the accuracy and reliability of attribution methods. High precision ensures the methods correctly identify actual threat actors while minimizing false positives. Researchers can drive continuous improvement in attribution technologies by emphasizing precision in evaluation metrics, leading to more accurate and trustworthy results.
\\
\\
In summary, the future of automated cyber threat attribution lies in integrating advanced AI techniques, developing real-time and explainable models, enhancing adversarial robustness, considering ethical and privacy implications, and benchmarking datasets and methods. Collaborative efforts within and across national borders will be essential for advancing the field and effectively combating the evolving landscape of cyber threats.

\section{Conclusion}
\label{sec:conclusion}
This survey provides an extensive review of the current landscape in cyber threat attribution and discusses the critical challenges and advancements that have shaped the field. 
The survey starts by exploring key artifacts that contribute to attribution and provides a taxonomy for scattered attribution artifacts. The taxonomy provides a standard way to understand and measure the possible attribution artifacts based on various qualitative measures. This survey also explores the available attribution datasets in the public domain and provides a classification of attribution datasets and their comparison. We further present the classification of available attribution methods based on the type of algorithm used and the dataset used for attribution. A comprehensive review of current literature on automated APT attribution is presented, along with critical comments and detailed discussions. Further, we analyze APT attribution’s current state and mention challenges toward advancing attribution methodologies. These comprehensive analysis of current literature led us to understand the future research directions in this domain and present the open research problems that researchers and practitioners can look into to advance the current state of APT attribution. 
By synthesizing these insights, this survey provides a comprehensive overview of the current state of APT attribution and offers valuable guidance for future research. We believe this work will inspire new advancements and practical solutions in threat attribution, ultimately fortifying our defenses against sophisticated cyber threats.  

\vspace{-0.5cm}
\section*{Acknowledgments}

This work has been partially supported by C3i (Cyber Security and Cyber Security for Cyber-Physical Systems) Innovation Hub at IIT Kanpur. Nanda Rani thanks the Prime Minister Research Fellowship (PMRF), Ministry of Education, Government of India, for the valuable financial support.
{

{\appendix
[Artifact analysis reasoning]
This section presents a rationale for analyzing the listed artifacts, based on the chosen evaluation measure.

\begin{table*}[]
    \centering
    \caption{Artifact analysis rationale glimpse}
    \label{tab:artifactcomparisonexplaination}
\begin{tabularx}{\textwidth}{p{1.6cm}p{2.9cm}p{2.9cm}p{2.9cm}p{2.9cm}p{2.9cm}}
        \hline
         \textbf{Artifact Name} & \textbf{Relevance} & \textbf{Integrity} & \textbf{Credibility} & \textbf{Timeliness} & \textbf{Accessibility} \\
         \hline
         Indicator of Compromise (IoCs) & IoCs like IP address, emails can be directly matched with any historical incident & Can be easily manipulated, spoofed, or reused by different threat actors & Often come from various reliable open-source intelligence and private sector reports & Available shortly after or even during an attack, allowing for quick detection and response & Often easy to collect and analyze, as they are usually widely shared and well-documented 
         \\ [10pt]
         C\&C Infrastructure & Evidence of using same infrastructure such as C\&C server can provide crucial link & Easy for attackers to disguise or alter without compromising their operations & Analyzed from network traffic analysis and collected from reliable open-source intelligence  & Can be identified and analyzed relatively quickly, but often requires ongoing monitoring and analysis & Requires specific tools and third-party database, and it can be challenging to analyze active C\&C servers 
         \\
         Cryptocurrency Transaction & These transactions are often anonymized and lack direct relevance to specific threat actor & The anonymity and potential use of mixers reduce reliability for attribution & Often derived from blockchain analysis, which is transparent and traceable & Often takes time due to the need to analyze blockchain data, and relevant information may only become clear after the transactions have been fully processed &  Tracking and analyzing these transactions can be complex due to the need for specialized tools and the anonymized nature of transactions
         \\
         Tactics, Techniques \& Procedures (TTPs) &  Represent the behavioral patterns of threat actors, which are often unique and difficult to fake & According to Pyramid of Pain, it is hard for attackers to alter without changing their operational methods &  Derived from detailed cybersecurity research and historical attack patterns, often published by reputable security firms & Generally identified through pattern analysis and research, and often emerge after analyzing multiple incidents & Accessible through threat intelligence reports and databases, but require in-depth analysis and understanding to link specific threat actor
         \\
         Malware \& Code Analysis & Reveal unique signatures and methods that are often tied to specific threat groups & Unique code signatures and methods are difficult to fake or alter significantly without detection & Conducted by skilled cybersecurity researchers and forensic experts, ensuring that the findings are trustworthy & Can be time-consuming, requiring in-depth investigation and analysis to uncover full spectrum of maliciousness &  Requires significant technical expertise and specialized tools to decompile and analyze code
         \\
         Toolchains & Can be indicative of specific groups, particularly when tied to known actor methodologies & Can be changed based on attacker group's preference and advancements in advanced tools & Often linked to direct observations by security professionals, including forensic investigations, providing high veracity & Requires detailed investigation, which can take time, but once identified, it can provide timely insights into ongoing campaigns &  Identifying and analyzing a threat actor's toolchain requires detailed investigation and correlation of multiple artifacts
         \\
         Language \& Writing Style & Less definitive on their own and may require corroboration with other artifacts & While unique, it can be imitated or altered, requiring careful consideration & Can be easily extracted from collected malware samples or received spam emails & Typically takes time, especially if it requires gathering sufficient samples for a conclusive analysis & Requires access to relevant samples and linguistic expertise 
         \\
         Public Claims and Hacker Forums & Not always relevant and reliable, may require validation & Potential for false claims, misinformation, or deliberate deception by actors & Often unverified and can be intentionally misleading & Can emerge quickly and sometimes delay depends on threat actor modus operandi & Relatively easy to access through OSINT, though its credibility varies
         \\
         Threat Intelligence Databases & Compile various artifacts and intelligence but rely on the quality and context of the data & Varies depending on the source and verification processes, making them moderately reliable & Curated by established cybersecurity firms, but can be updated by non-credible individuals & Updated regularly, but there can be a delay between the detection of new threats and the information inclusion in the database &  Designed to be easily accessible to analysts, providing structured information that is easy to query and analyze
         \\

         Geopolitical & Can offer valuable insights into potential motivations and likely actors, but needs to be combined with other relevant evidences & Interpretative but may be influenced by biases or incomplete information, still provides valuable context in some cases & Can be reliable, but may also be influenced by political biases or incomplete information & Requires a broad and often slow-gathering of information &  Requires a good understanding of current geopolitical contexts and access to relevant intelligence reports
         \\
         Victimology & Highly relevant, as specific threat actors often target certain industries or regions, providing clues to their identity &  Relies on patterns that could potentially be aid if attackers intentionally target victims & Often based on reported incidents, patterns and data collected from victims & Can provide timely insights once sufficient data on targeted entities is gathered & Requires detailed analysis of attack patterns and affected entities, which can be resource-intensive
         \\
         Post-Incident Communication & May offer insights into the actors involved, but it is less direct and may require interpretation &  May provide reliable insights, but may also be subject to manipulation or deception & Usually comes from victim's conversation with threat actor & Typically emerge with a delay in engaging threat actor in communication &  Can be accessible through monitoring communication channels, but requires active intelligence gathering
         \\
         Time Zone Analysis & Can suggest the likely origin of the threat actor, it is a circumstantial piece of evidence & Circumstantial and can be easily misrepresented by attackers using tools to obfuscate their actual location &  Often based on generally reliable metadata and logs & Conducted quickly by leveraging metadata from logs or other artifacts, providing timely insights into potential origins &  Relatively straightforward and can often be accessible from metadata of logs and other time-stamped artifacts
         \\
         \hline
    \end{tabularx}
    \end{table*}
 }

\end{document}